\documentclass[twocolumn,superscriptaddress,aps,pra,nofootinbib,floatfix,10pt]{revtex4-2}
\usepackage{amssymb}
\usepackage{amsmath}
\usepackage{graphicx}% Include figure files
\usepackage{dcolumn}% Align table columns on decimal point
\usepackage{bm}% bold math
\usepackage{dsfont}
\usepackage[breaklinks=true,colorlinks,citecolor=blue,linkcolor=blue,urlcolor=blue]{hyperref}% add hypertext capabilities
\usepackage[caption=false]{subfig}
\usepackage{physics}
\usepackage{mleftright}
\usepackage{bbm}
\usepackage{soul}
\usepackage{enumerate}
\usepackage{tikz}
\usetikzlibrary{shapes, arrows, positioning}
\usepackage{amsmath}
\newcommand{\be}{\begin{equation}}
\newcommand{\ee}{\end{equation}}
\newcommand{\bea}{\begin{eqnarray}}
\newcommand{\eea}{\end{eqnarray}}

\newcommand{\figref}[1]{\mbox{Fig.~\ref{#1}}}

\newcommand{\secref}[1]{\mbox{Sec.~\ref{#1}}}

\newcommand{\appref}[1]{\mbox{Appendix~\ref{#1}}}
\renewcommand{\eqref}[1]{\mbox{Eq.~(\ref{#1})}}

\newcommand{\figpanel}[2]{Fig.~\hyperref[#1]{\ref*{#1}(#2)}}

\usepackage{tikz,xcolor}
\definecolor{lime}{HTML}{A6CE39}
\DeclareRobustCommand{\orcidicon}{%
	\begin{tikzpicture}
	\draw[lime, fill=lime] (0,0) 
	circle [radius=0.16] 
	node[white] {{\fontfamily{qag}\selectfont \tiny ID}};
	\draw[white, fill=white] (-0.0625,0.095) 
	circle [radius=0.007];
	\end{tikzpicture}
	\hspace{-2mm}
}
\foreach \x in {A, ..., Z}{%
	\expandafter\xdef\csname orcid\x\endcsname{\noexpand\href{https://orcid.org/\csname orcidauthor\x\endcsname}{\noexpand\orcidicon}}
}

\begin{document}
%%%%%%%%%%%%%%%%%%%%%%%%%%%%%%%%%%%%%%%%%%%%%%%%%%%%%%%%%%%%%%%%%%%%%%
%%%%%%%%%%%%%%%%%%%%%%%%%%%%%%%%%%%%%%%%%%%%%%%%%%%%%%%%%%%%%%%%%%%%%%
\title{An asymptotic field approach for the control of dipole emission in integrated structures} 

\date{\today}
\author{Vincenzo Macr\`{i}\orcidE{}}
\email{vincenzo.macri@unipv.it}
\affiliation{{Dipartimento di Fisica "A. Volta", Università di Pavia, Via Bassi 6, 27100 Pavia, Italy}}

\author{Alice Viola\orcidA{}}
%\email{alice.viola01@universitadipavia.it}
\affiliation{{Dipartimento di Fisica "A. Volta", Università di Pavia, Via Bassi 6, 27100 Pavia, Italy}}

\author{Marco Liscidini}
\email{marco.liscidini@unipv.it}
\affiliation{{Dipartimento di Fisica "A. Volta", Università di Pavia, Via Bassi 6, 27100 Pavia, Italy}}

\begin{abstract}
We present a general framework to model spontaneous emission in integrated photonic structures by exploiting quantization of the electromagnetic field in terms of asymptotic in/out modes. This approach allows for an efficient and physically meaningful  calculation of the emission rate into each radiative channel of an arbitrary structure, without relying on approximations such as Lorentzian lineshapes or point-like system-bath coupling. We show that with this approach one can recover well-known results for dipole emission in waveguides or ring resonators, and that such results can be easily extended to include the effect of backscattering. Finally, as an application, we design a tunable integrated single-photon source that enables full control over both the emission rate and output mode. This flexibility makes our method particularly well-suited for the design and analysis of integrated single-photon sources in various material platforms.
\end{abstract}
\maketitle

\section{Introduction}
Spontaneous emission (SE) is a notable example of quantum electrodynamics, in which vacuum fluctuations lead to the emission of a photon from an excited state population, even in the absence of coherence \cite{Fermi1932,weisskopf1997berechnung}. Comprehending and manipulating SE is fundamentally important across various domains in nanophotonics \cite{novotny2012}, including the investigation of nano-lasers \cite{chow2014,deng2021}, active fibers \cite{Tromborg2001}, exceptional points \cite{Lin2016,pick2017}, and coupled loss-gain systems \cite{peng2014}. In particular, single-photon emission represents a crucial resource for the deterministic generation of non-Gaussian quantum states of light, which are central to many applications in quantum information processing, quantum communication \cite{Genoni2007,smith2011,lee2019}, and quantum metrology \cite{Huang2018,Walschaers2021,Xu2022}. 

Achieving high-quality single-photon sources is a significant challenge, and current efforts typically rely on embedding quantum dots within micropillars, which provide efficient photon extraction and high emission rates \cite{Senellart2017}, hybrid integration of multiple InAs/InP microchiplets containing high-brightness quantum dot single photon emitters into advanced silicon-on-insulator photonic integrated circuits \cite{larocque2024tunable}, exploiting defect-based emitters (e.g. nitrogen-vacancy (N-V) centres in diamond) integrated into photonic crystal structures \cite{Ruf2019} or in a cavity composed of two nanospheres with a trapped quantum emitter \cite{Groiseau2024}. Despite these advancements, developing sources that meet spectral purity, indistinguishability, high rate, and nearly unitary collection efficiency, which are essential to practical applications, continues to be an area of active research. To this end, an important aspect is the ability to leverage integrated photonic complexity by modelling light emission in complex photonic structures in such a way that system parameters can be directly linked to the properties of the emitted light. 

Here, we investigate the use of an asymptotic-in/out field formalism \cite{Liscidini2012}, initially introduced to model the nonlinear light-matter interaction, to describe dipole emission in a generic photonic structure. This approach allows us to easily calculate the emission rate for any radiative channel of the structure by quantizing the electromagnetic field starting from stationary solutions of Maxwell’s equations. Within this framework, the complexity of the structure geometry can then be taken into account analytically or numerically, providing easy integration with numerical tools for modeling nanophotonic devices. Furthermore, our strategy remains applicable even when the resonant field enhancement arises from non-Lorentzian resonances. This extends beyond the commonly used point-coupling approximation typically assumed in approaches based on Gardiner–Collett-type Hamiltonians for modeling light–matter interactions. As a practical application, we design a single-photon source based on SE in a ring resonator, enabling deterministic control over the dipole emission. Thanks to our approach, we are able to derive explicit analytical results in terms of asymptotic-in and -out fields, demonstrating that the proposed design allows direct control of the emission rate and optical mode of the emitted photons. 

The paper is structured as follows. In \secref{model}, we derive a simple yet general expression for calculating the emission rate of a point dipole based on the asymptotic fields of an arbitrary structure. We demonstrate that this approach can reproduce well-known results in simple geometries, such as a straight waveguide and a ring resonator. In \secref{General}, we apply our formulation to the design and simulation of an integrated structure capable of achieving deterministic single-photon emission in a well-defined mode. We show that our description of dipole emission provides insight into the underlying physics governing the operation of the device. In \secref{conclu} we present our conclusions. Some details are left for the appendices: starting from the Fermi’s golden rule, in \appref{appendixa} we found a general formula for the total dipole emission rate, in \appref{Appendix1} we derive the correct generic form for the scattering part of the asymptotic field in the case of a waveguide coupled to a ring resonator, while in \appref{Appendix_backscattering2} we repeat the same derivation in presence of a lumped scatter, and finally, in \appref{Appendix2} we extend the method employed to a more general geometry. 

\section{Dipole emission rate with an asymptotic field approach }\label{model}
In this section, we address the problem of calculating the spontaneous emission rate of a quantum emitter coupled to a generic photonic structure. The electromagnetic environment, such as waveguides, cavities, or more complex nanophotonic systems, significantly influences the emission properties of the emitter by modifying the photonic mode structure and the local density of optical states. To account for these effects within a general framework, we rely on perturbation theory and, in particular, on Fermi’s Golden Rule, which provides a powerful method for calculating the transition rate $\Gamma$ from an initial quantum state $|i\rangle$ to one or more final states $|f\rangle$ as \cite{dirac1927,fermi1950,sakurai2020}:
\begin{equation}
\Gamma = \frac{2\pi}{\hbar} \sum_f \left| \langle f | \hat{H}_{I} | i \rangle \right|^2 \delta(E_f - E_i).
\label{fermigoldenrule}
\end{equation}
Here, $E_i$ and $E_f$ are the energies of the initial and final states, respectively, and $\hat{H}_{I}$ is the interaction Hamiltonian responsible for the transition. 
\begin{figure}[ht!]
	\centering
 \includegraphics[width=1\linewidth]{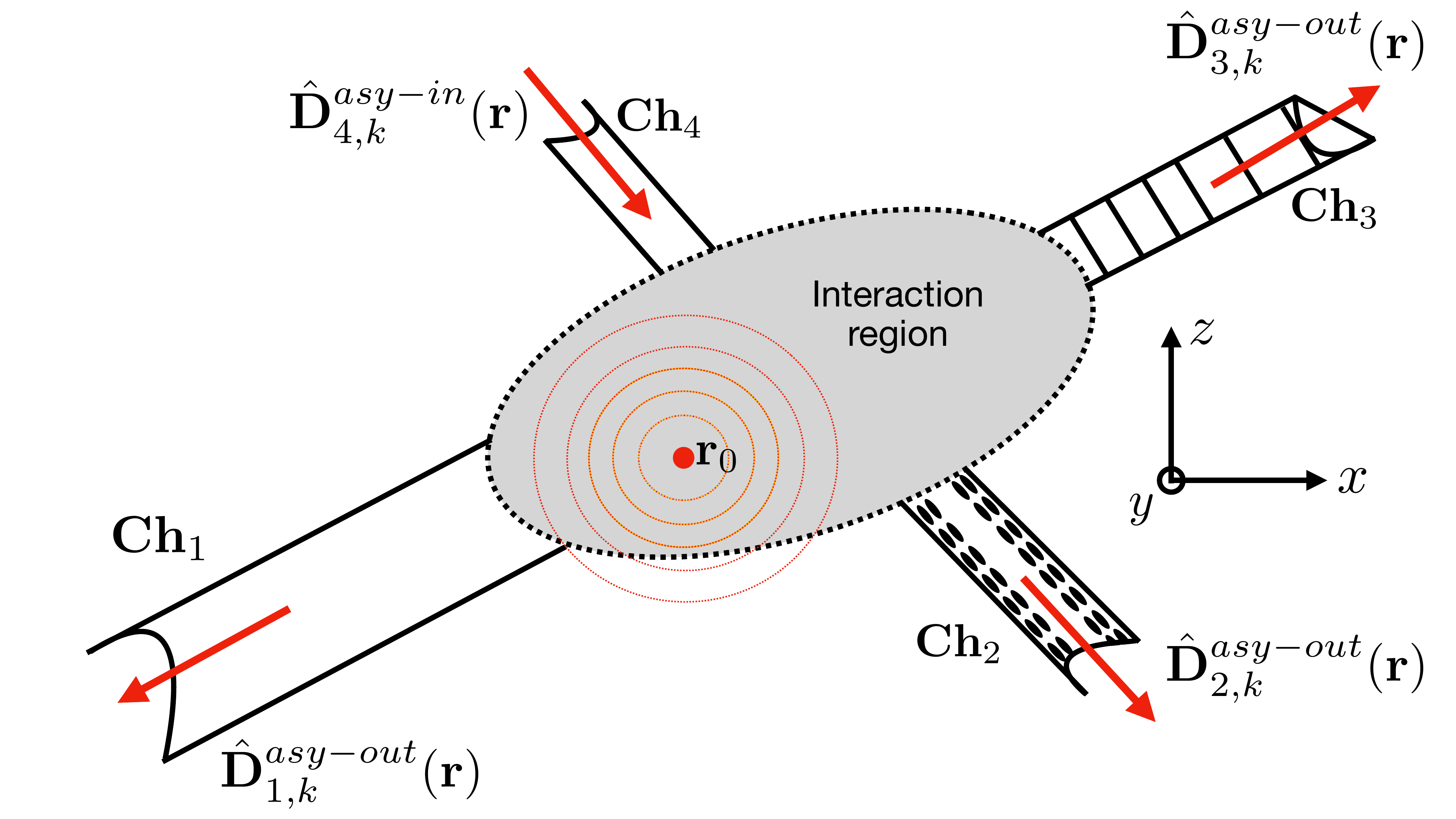}
	\caption{A sketch of the kind of structure of interest. The red spot represents the dipole inside the interaction region.}
	\label{Figure1}
\end{figure}

In this work, we shall focus on a point-like dipole interacting with the electromagnetic field, where \( \hat{H}_{I} \) is given by
\begin{equation}\label{interactionI}
\hat{H}_{I} = \int_{\textbf{r}}\textrm{d} \textbf{r}~\hat{\mathbf{p}}\cdot \hat{\textbf{E}}(\textbf{r}) \delta(\textbf{r}-\textbf{r}_0)=\hat{\mathbf{p}}\cdot \hat{\textbf{E}}(\textbf{r}_0) \;,
\end{equation}
where $\hat{\mathbf{p}}=\mathbf{p}\hat{\sigma}_x$ is the usual dipole operator in terms of the  Pauli $x-$matrix, and $\textbf{r}_0$ is the position of the dipole inside a generic structure in the laboratory frame (see \figref{Figure1}), and $\hat{\textbf{E}}(\textbf{r})$ is the electric field operator.

Since our goal is the calculation of the emission rate far from the structure, we adopt the asymptotic-in/out fields approach \cite{Liscidini2012} and write the electric field operator as 
\begin{equation}\label{interactionII}
\hat{\mathbf{E}}(\textbf{r})=\sum_{N}\int_0^\infty dk\sqrt{\frac{\hbar\omega_{N,k}}{2}}\hat{a}_{N,k} \frac{\mathbf{D}^\text{asy-in(out)}_{N,k}(\mathbf{r})}{\varepsilon_0 n^2(\mathbf{r})}+ \text{H.c}.
\end{equation}
where $\mathbf{D}^{\text{asy-in(out)}}_{N,k}(\mathbf{r})$ is the asymptotic-in (out) field associated with the $N$-th channel of the structure, which we assume, for simplicity, to be single-mode. Here, $k$ is the wavevector, $\omega_{N,k}$ the corresponding angular frequency, $n(\mathbf{r})$ the refractive index, and $\varepsilon_0$ the vacuum permittivity. The operators $\hat{a}_{N,k}$ and $\hat{a}^{\dagger}_{N,k}$ represent the annihilation and creation operators, respectively, and satisfy the usual commutation relation $[\hat{a}_{N,k}, \hat{a}^{\dagger}_{N’,k’}] = \delta_{N,N’} \delta(k’ - k)$ \cite{Liscidini2012,quesada2022}. We assume that no electromagnetic field modes are bound to the interaction region. Finally, we consider the case of transparent media, where the asymptotic-out field is related to the asymptotic-in field by $\mathbf{D}^{\text{asy-out}}_{N,k}(\mathbf{r}) = \left[\mathbf{D}^{\text{asy-in}}_{N,k}(\mathbf{r})\right]^*$.

For simplicity, we shall consider a dipole emitting in a very narrow frequency range centered at $\omega_0$ such that,  for each channel, we can expand the dispersion relation at the first order as $k(\omega)=k_{0,N}+\frac{1}{{v}_{g_{N}}}(\omega-\omega_0)$, with $k_{0,N}$ the wavevector at $\omega_0$, ${v}_{g_{N}}$ the group velocity. Under this hypothesis, one obtains (see \appref{appendixa}) that the total dipole emission rate can be written as  
\begin{eqnarray}
\Gamma = \frac{\pi \omega_0 }{\varepsilon_0^2 n^4(\mathbf{r_0}) \hbar} \sum_{N} \frac{1}{v_{g_{N}}}\left|\mathbf{\hat{p}}\cdot\mathbf{D}^\text{asy-out}_{N,k_{0,N}}(\mathbf{r}_0)\right|^2. 
\label{fermigoldenrule2}
\end{eqnarray}
From this expression, one can observe that  the quantization of the electric field in terms of asymptotic fields allows one to immediately identify the emission rate in each channel $N$ 
\begin{equation}\label{fermi_golden}
\Gamma_N = \frac{\pi \omega_0 }{\varepsilon_0^2 n^4(\mathbf{r_0})\hbar} \frac{1}{v_{g_{N}}}\left|\mathbf{\hat{p}}\cdot\mathbf{D}^\mathrm{asy-out}_{N,k_{0,N}}(\mathbf{r}_0)\right|^2.
\end{equation}
In the following, we will use this expression to calculate the dipole emission rate in simple structures, both to demonstrate the flexibility of the approach and to highlight key physical aspects of dipole emission in a photonic environment that are naturally captured by our formalism. In doing this, we will evaluate the emission rate starting from the asymptotic-in fields rather than the asymptotic-out by exploiting the aforementioned relation $\mathbf{D}^{\text{asy-out}}_{N,k_{0,N}}(\mathbf{r}) = \left[\mathbf{D}^{\text{asy-in}}_{N,k_{0,N}}(\mathbf{r})\right]^*$ \cite{Liscidini2012}. More in general, it is up to the reader to decide which field (in or out) is more convenient to use, depending on the system under consideration and whether the fields are evaluated analytically or numerically.

\subsection{Dipole emission in a waveguide}
\begin{figure*}[t!]
	\centering
 \includegraphics[width=\linewidth]{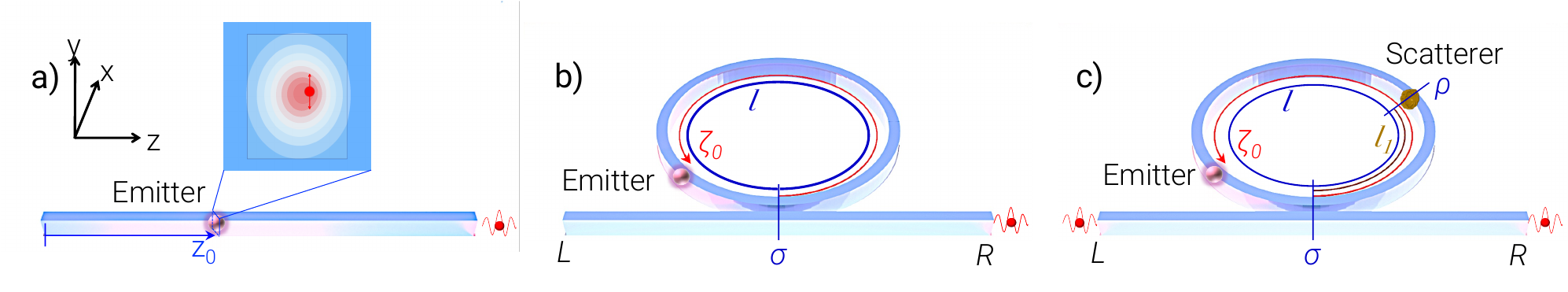}
	\caption{Sketch of the structures for enhanced dipole emission; (a) straight, single-mode waveguide. The red dot indicates the dipole, located at $\mathbf{r}_0 = (x_0, y_0, z_0)$. The two radiative channels of interest are called  L (from the left) and R (from the right). The inset depicts the waveguide cross-section, with the dipole (the red dot with vertical arrows) emitting into the guided mode. (b) Ring resonator. The dipole is located at a longitudinal position $\zeta_0$ along the ring. The resonator supports both clockwise and counterclockwise modes. (c) Ring resonator with a localized scatterer, indicated as a brown mark at position $ l_1$ along the ring. The scatterer introduces a coupling between the clockwise and counterclockwise modes, which affects both the total emission rate and the direction of the emitted photons.}
	\label{StructuresFigure}
\end{figure*}
We start by considering the simple case of a dipole inside a lossless single-mode waveguide, and we shall assume that the dipole emission is significant only in the guided modes of the structure. We consider the reference frame shown in \figpanel{StructuresFigure}{a} with the dipole position being  $r_0=(x_0,y_0,z_0)$.
For such a simple structure, the asymptotic fields are nothing more than the solutions of Maxwell equations corresponding to light incoming from left (L) or right (R) and propagating in the waveguide, with the displacement field completely determined by the transverse field profile $\mathbf{d}(x,y)$ (which is normalized as in \cite{Yang2008, Liscidini2012}) and by the wavevector component $k$ (or -$k$) along the propagation direction $z$. At the dipole position, one can write
\begin{eqnarray}
   \mathbf{D}_{L,k_0}^\text{asy-in}(\mathbf{r}_0)&=&\frac{\mathbf{d}(x_0,y_0)}{\sqrt{2 \pi}}  e^{ik_0 z_0}   \\
     \mathbf{D}_{R,k_0}^\text{asy-in}(\mathbf{r}_0)&=&\frac{\mathbf{d}(x_0,y_0)}{\sqrt{2 \pi}}   e^{-ik_0 z_0},\quad
    \label{waveguideAsy-Filed}
\end{eqnarray}
where $k_0$ is the wavevector at the dipole emission frequency. 

By considering a dipole oriented along the displacement field, from \eqref{fermigoldenrule2} we obtain the total emission rate in the waveguide modes (L+R) as
\begin{eqnarray}
\Gamma_{\text{wg}} &=& \frac{p^2 \omega_0}{\varepsilon_0 n^2(\mathbf{r_0}) \hbar v_g} \frac{d^2(x_0,y_0)}{\varepsilon_0 n^2(\mathbf{r_0})}  \nonumber \\
&=& \frac{p^2 \omega_0}{\varepsilon_0 n^2(\mathbf{r_0}) \hbar v_g}\frac{1}{ A_{\rm eff}(\mathbf{r_0})}  \;,
\label{GammaWaveGuide}
\end{eqnarray}
where we introduce the mode effective area associated with the dipole interaction
\begin{align}
A_{\rm eff}(\mathbf{r_0}) = \frac{\varepsilon_0 n^2(\textbf{r}_0)}{d^2(x_0,y_0)}
\label{areaeff1}\; .
\end{align}
We note that when the dipole is placed at the position of maximum energy density, the effective area reaches its minimum value, given by $
\bar{A}_{\rm eff} = \frac{\int \mathrm{d} \mathbf{r}\, \varepsilon(\mathbf{r})|\mathbf{E}(\mathbf{r})|^2}{\max[\varepsilon(\mathbf{r}) |\mathbf{E}(\mathbf{r})|^2]},$
which is commonly referred to as the mode effective area.

It is interesting to compare the result we achieved for the dipole emission rate inside a waveguide with the usual emission rate of a randomly oriented dipole in an isotropic dielectric medium with refractive index $n$ (e.g. see \cite{Kien2016}), given by $\Gamma_0= p^2 \omega_0^3/ 3 \varepsilon_0 \hbar \pi c^3$. One obtains 
\begin{align}\label{purcell_waveguide}
\frac{\Gamma_{\text{wg}}}{n\Gamma_0}=\frac{3}{4\pi}\frac{n_g}{n}\left(\frac{\lambda_0}{n}\right)^2\frac{1}{A_{\rm eff}(\mathbf{r_0})}\;,
\end{align}
where $\lambda_0=2\pi c/\omega_0$ which demonstrates that the emission rate probability in a waveguide increases with the inverse of the effective area, expressed in units of the wavelength within the waveguide material. Such a result can be regarded as the waveguide analog of the more well-known Purcell factor for dipole emission in a resonator. A second important aspect highlighted in \eqref{purcell_waveguide} is the role of the group index: specifically, the emission rate increases with the inverse of the group velocity. This effect is well known in the context of photonic crystal waveguides, where very low group velocities (corresponding to group indices $n_g$ as high as \cite{bogaerts2012,OFaolain:10}) can be achieved thanks to the high degree of flexibility in dispersion engineering.

\subsection{Dipole emission in a ring resonator}
\label{RingEmission}
We now consider dipole emission in a resonant system, specifically, a ring resonator coupled to a waveguide, as illustrated in \figpanel{StructuresFigure}{b}.  
In \appref{Appendix1}, we show how to derive the asymptotic-in fields $\mathbf{D}^\text{asy-in (ring)}_{N,k}(\mathbf{r}_0)$ in such a structure by using a scattering matrix model. In doing so, we distinguish the clockwise and counterclockwise modes, which correspond to the asymptotic fields entering from the right (R) and left (L) ports, respectively. One has 
\begin{eqnarray}
   \mathbf{D}_{L,k_0}^\text{asy-in (ring)}(\mathbf{r}_0)&=& \frac{i \kappa}{1-\sigma e^{i \delta_0}} \frac{\mathbf{d}(\textbf{r}_{\perp,0},\zeta_0)}{\sqrt{2\pi}}  e^{i \tilde{\delta}_0},  \\
     \mathbf{D}_{R,k_0}^\text{asy-in (ring)}(\mathbf{r}_0)&=& \frac{i \kappa}{1-\sigma e^{ i \delta_0}} \frac{\mathbf{d}(\textbf{r}_{\perp,0},\zeta_0)}{\sqrt{2\pi}}  e^{i (k_0l - \tilde{\delta}_0)},\;
    \label{ringAsy-Filed}
\end{eqnarray}
where $\sigma$ is the self-coupling coefficient between the waveguide and the ring, and $\kappa = \sqrt{1 - \sigma^2}$ (with $\sigma, \kappa \in \mathbb{R}
$). The function $\mathbf{d}(\mathbf{r}_{\perp,0}, \zeta_0)$ represents the mode field profile evaluated at the dipole position $(\mathbf{r}_{\perp,0}, \zeta_0)$, where $\mathbf{r}_{\perp,0}$ and $\zeta_0$ denote the transverse and longitudinal coordinates, respectively, within the ring. Finally, $l$ is the resonator length, $\delta_0 = k_0 l$, and $\tilde{\delta}_0 = k_0 \zeta_0$.

From \eqref{fermigoldenrule2}, one can readily obtain the total dipole emission rate, given by the sum of the emission into the clockwise and counterclockwise modes, as
\begin{align}
\Gamma_\mathrm{ring} =& \frac{p^2 \omega_0}{\varepsilon_0 n^2(\mathbf{r_0}) \hbar v_g} \frac{ d^2(\textbf{r}_{\perp,0},\zeta_0)}{\varepsilon_0 n^2(\textbf{r}_{\perp,0},\zeta_0)}    \frac{ \kappa^2}{1+\sigma^2 - 2\sigma\cos{\delta_0}} \nonumber\\
=& \Gamma_{\text{wg}} \frac{ \kappa^2}{1+\sigma^2 - 2\sigma\cos{\delta_0}} \;,
\label{fermigoldenrule3}
\end{align}
where $A_{\rm eff}(\mathbf{r_0}) = \frac{\varepsilon_0 n^2(\textbf{r}_{\perp,0},\zeta_0)}{d^2(\textbf{r}_{\perp,0},\zeta_0)}$ is the effective area of guided mode in the resonator waveguide evaluated at the dipole position. 

\eqref{fermigoldenrule3} naturally follows from our formalism and highlights the distinct roles of transverse and longitudinal confinement. The transverse field confinement, as expected, yields the same enhancement as in a simple waveguide. In contrast, the spatio-temporal longitudinal confinement, associated with the resonant response of the ring, provides an additional increase in the emission rate. Assuming the dipole is resonantly coupled to the cavity and considering the high-finesse limit ($\sigma \approx 1$), when the ring exhibits a resonant Lorentzian response, the expression can be further simplified as
\begin{equation}\label{fermigoldenrule6}
\Gamma_\mathrm{ring}^{\sigma \approx 1}  =\Gamma_{\text{wg}}  \frac{2}{1-\sigma}=2 n\Gamma_0 \frac{3 }{4 \pi^2} \biggr{(}\frac{\lambda_0}{n}\biggr{)}^3 \frac{Q}{V_{\rm eff}} \;,  
\end{equation}
where $Q$ is the resonance quality factor. This expression agrees with the well-known Purcell result \cite{purcell1995}, which describes the enhancement of the dipole emission rate in resonant systems. Here, we introduced the effective volume $V_\mathrm{eff} = \bar{A}_\mathrm{eff} l$, and we used the fact that, in the high-finesse limit, $2/(1 - \sigma) = 4 v_g Q / \omega_0 l$ (see \cite{onodera2016}), with $v_g = c / n_g$ being the group velocity. The factor of 2 in front of the second term accounts for emission into both the clockwise and counterclockwise modes.

We emphasize that the Purcell factor in \eqref{fermigoldenrule6} is derived—as is customary—under specific assumptions: high finesse, single-mode emission, a dipole placed at the maximum of the energy density, and resonant coupling between the dipole and the resonator \cite{rosenfeld2003,bogaerts2012,fait2021,flaagan2022}. In contrast, our formalism yields more general results. For instance, \eqref{fermigoldenrule3} remains valid even when the dipole is not resonantly coupled to the ring or in the low-finesse regime, where the enhancement of the emission rate arises primarily from transverse field confinement. 

\subsection{Dipole emission in a resonator including backscattering}

To further highlight the flexibility of our approach, we now extend the results of \secref{RingEmission} to include the effect of backscattering inside the ring resonator.  
Backscattering can arise from various imperfections in the ring, such as edge roughness \cite{poulton2006,Morichetti2010,bogaerts2012,lee2001}, fabrication defects \cite{smith2000,Hughes2005}, or even the dipole itself \cite{li2016}.  Here, we model backscattering as originating from a single localized scatterer — see \figpanel{StructuresFigure}{c} — that affects light propagation in the ring. The scatterer is characterized by a real reflection coefficient $\rho$ and a transmission coefficient $\tau = \sqrt{1 - \rho^2}$ \cite{Little_1997}, where $|\rho|^2$ represents the probability of backscattering. This approach can also be extended to include multiple scatterers, but such generalizations are beyond the scope of this work.

Unlike the simpler case discussed earlier, the clockwise and counterclockwise modes are now coupled. As a result, within the ring resonator, the asymptotic fields for the left and right ports are general superpositions of counter-propagating modes (see \appref{Appendix_backscattering2}). We can therefore write:
\begin{eqnarray}
   \mathbf{D}_{L,k_0}^\text{asy-in (ring)}(\mathbf{r}_0)&=& f_{k_0}^{(L)}(\zeta_0) \frac{\mathbf{d}(\textbf{r}_{\perp,0},\zeta_0)}{\sqrt{2\pi}},   \\
     \mathbf{D}_{R,k_0}^\text{asy-in (ring)}(\mathbf{r}_0)&=& f_{k_0}^{(R)}(\zeta_0) \frac{\mathbf{d}(\textbf{r}_{\perp,0},\zeta_0)}{\sqrt{2\pi}},
    \label{asymptotic_backS}
\end{eqnarray}
where $f_{k_0}^{(L, R)}(\zeta_0)$ describe the field enhancements at the dipole location $(\mathbf{r}_{\perp,0}, \zeta_0)$, associated with the asymptotic-{in} %out
fields for the left (L) and right (R) ports. These quantities are given explicitly in \eqref{Fielda_Back_system} and \eqref{Fieldb_back_system}.

It follows that the total emission rate $\Gamma_{\text{rb}}$, in the presence of backscattering, can be written as  
\begin{align}
    \Gamma_{\text{rb}} =& \frac{\Gamma_{\text{wg}}}{2} \biggl( \left|f_{k_0}^{(L)}(\zeta_0)\right|^2  + \left|f_{k_0}^{(R)}(\zeta_0)\right|^2 \biggr) \nonumber \\
    =& \frac{\Gamma_{\text{wg}}}{2} \kappa^2 \frac{(1 + \sigma^2 - 2\tau \sigma e^{i \delta_0})(1 - i \rho e^{-2 i \Delta}) + \text{H.c.}}{\left|\rho^2 + (\tau - \sigma e^{i \delta_0})^2 \right|^2}\;.
\label{golden_rule_backscattering2}
\end{align}
Here, $\Delta = \tilde{\delta}_0 - \delta_1$, with $\delta_1 = k_0 l_1$, where $l_1$ denotes the position of the scatterer (see \figpanel{StructuresFigure}{c}). The parameter $\Delta$ represents the phase mismatch due to the difference in longitudinal position between the dipole and the scatterer.
This phase mismatch plays a crucial role in determining the impact of backscattering on the dipole emission when $\rho \neq 0$.

Unlike the case with no backscattering, in general the emission rates $\Gamma_{\text{rb,L}}$ and $\Gamma_{\text{rb,R}}$ from the left and right ports are different, and can be calculated directly from Eqs.~(\ref{Fielda_Back_system}-\ref{Fieldb_back_system}) by considering the emission into each asymptotic field mode. These quantities, normalized to the emission rate in a simple waveguide, can be written as it follows 
\begin{align}
   \frac{\Gamma_{\text{rb,L}}}{\Gamma_{\text{wg}}}  = &   \frac{\biggl[\kappa^2(\sigma^2 -  \tau \sigma e^{i \delta_0})(1 - i \rho e^{-2 i \Delta}) + \text{H.c.} \biggr] + \kappa^4 \tau^2}{2 \left|\rho^2 + (\tau - \sigma e^{i \delta_0})^2 \right|^2} \label{golden_rule_backscattering3a} \\
 \frac{\Gamma_{\text{rb,R}}}{\Gamma_{\text{wg}}}  = &  \frac{\biggl[\kappa^2(1 -  \tau \sigma e^{i \delta_0})(1 - i \rho e^{-2 i \Delta}) + \text{H.c.} \biggr] - \kappa^4 \tau^2}{2 \left|\rho^2 + (\tau - \sigma e^{i \delta_0})^2 \right|^2} \;.
\label{golden_rule_backscattering3}
\end{align}

From Eqs.~(\ref{golden_rule_backscattering2}-\ref{golden_rule_backscattering3}), it is clear that the parameter $\Delta$ determines both the total emission rate and the distribution of emitted photons between the two ports of the structure. When backscattering arises from random inhomogeneities in the fabrication process, $\Delta$ cannot be controlled. However, one can envision using nanofabrication to introduce intentionally designed reflectors in specific regions of the resonator structure (e.g., see \cite{arabi_2011}). In such cases, it becomes possible, in principle, to have some control over the emission properties.  It is worth noting that, due to the resonant field enhancement within the cavity, even moderate backscattering can produce a significant modification of both the emission rate and the directional distribution of the emitted light.

For this reason, it is useful to consider the case in which the dipole is resonantly coupled to the ring, i.e., when $\delta_0 = 2m\pi$ with $m \in \mathbb{N}$. In this case, the expressions in Eqs.~(\ref{golden_rule_backscattering2}-\ref{golden_rule_backscattering3}) simplify to  
\begin{align}\label{golden_rule_backscattering4}
\Gamma_{\text{rb}} = {\Gamma_{\text{wg}}}\kappa^2 \frac{1 - \rho \sin{2\Delta}}{\rho^2 + (\tau - \sigma)^2} \;,
\end{align}
and  
\begin{align}\label{golden_rule_backscattering5}
   \frac{\Gamma_{\text{rb, L}}}{\Gamma_{\text{rb}}}  = & \frac{2 \sigma(\sigma - \tau) (1 - \rho \sin{2\Delta}) + \kappa^2 \tau^2}{2 (1 - \rho \sin{2\Delta})(\rho^2 + (\tau - \sigma)^2)}\;, \\
 \frac{\Gamma_{\text{rb, R}}}{\Gamma_{\text{rb}}}  = & \frac{2 (1 - \tau \sigma) (1 - \rho \sin{2\Delta}) - \kappa^2 \tau^2}{2 (1 - \rho \sin{2\Delta})(\rho^2 + (\tau - \sigma)^2)}\;. \label{golden_rule_backscattering5b}
\end{align}
Here, \eqref{golden_rule_backscattering5} and \eqref{golden_rule_backscattering5b} represent the probabilities that a photon emitted into the ring exits from the left and right ports, respectively.
\begin{figure}[t!]
	\centering
 \includegraphics[width=1.1\linewidth]{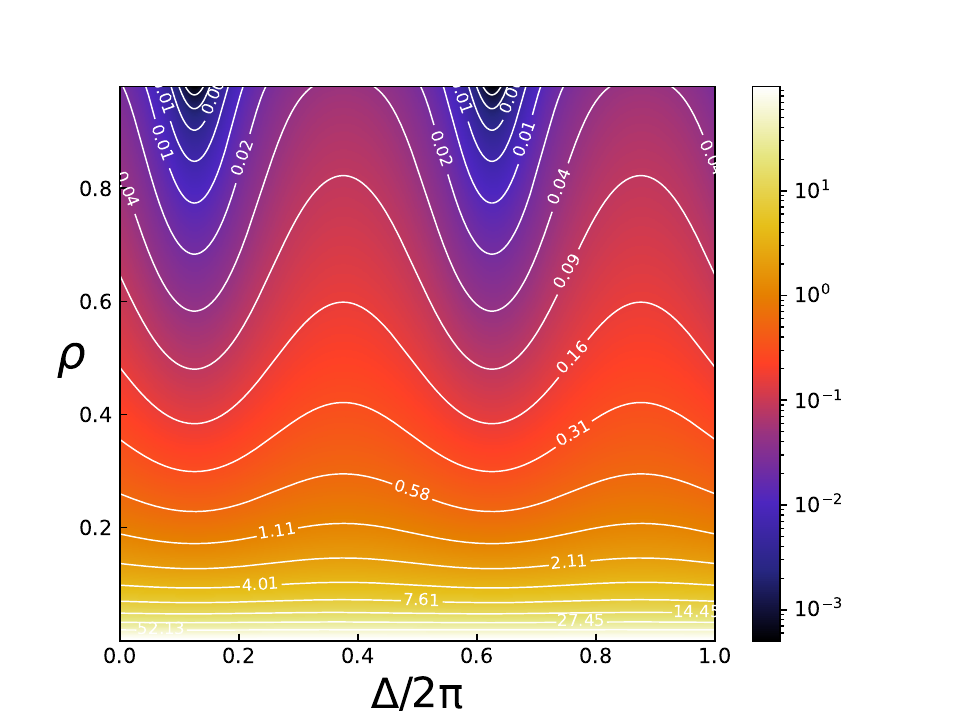}
	\caption{Contour plot  of the normalized dipole emission rate $\Gamma_{\text{rb}}/\Gamma_{\text{wg}}$ in log scale as function of backscattering phase $\Delta$, and the reflectivity $\rho$. In particular, the emission can be completely suppressed when $\Delta = \pi/4 + m\pi$, where $m \in \mathbb{N}$. As expected, when $\rho\rightarrow 0$, one obtains the result for the backscatter-free ring. Parameters used: $\lambda= 630 nm$, length $l=300\pi\lambda_0=93.7$ $\mu$m, and self-coupling parameter $\sigma=0.98$.}
	\label{ContourPLot}
\end{figure}
\begin{figure}[ht!]
	\centering
 \includegraphics[width=\linewidth]{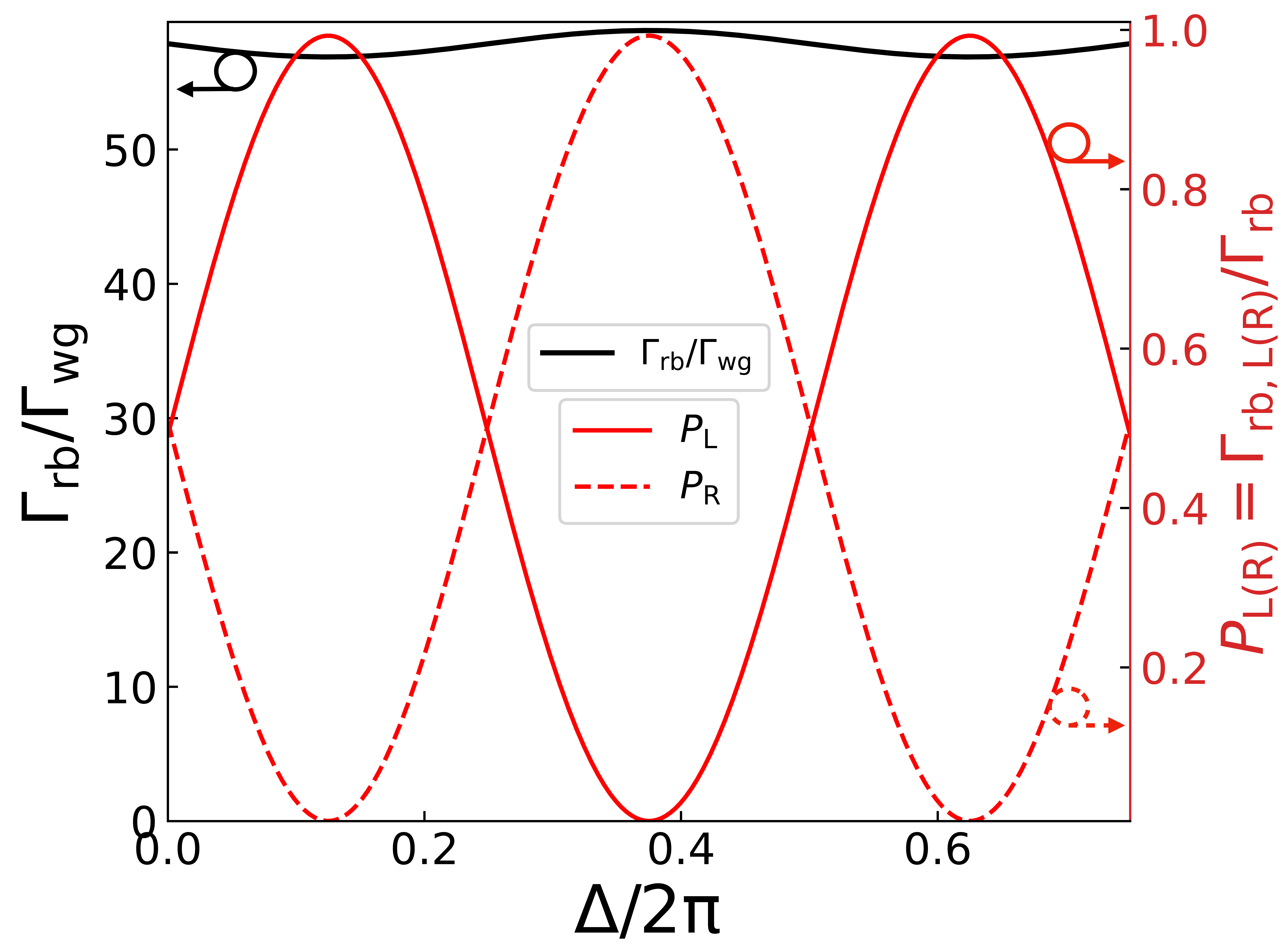}
	\caption{The normalized dipole emission rate, $\Gamma_{\text{rb}}/\Gamma_{\text{wg}}$ (black solid line), and the normalized output probability,  $P_{L(R)}=\Gamma_\mathrm{rb, L(R)}/\Gamma_\mathrm{rb}$ (red solid and dashed line, respectively), are shown as functions of the backscattering phase $\Delta$. Notably, while the overall dipole emission enhancement remains nearly constant, the probability of photon emission from the left or right port can be tuned by adjusting the dipole position within the ring.
    The parameters used in the simulation are: $\lambda_0 = 630\;\text{nm}$, length $l=300\pi\lambda_0=93.7$ $\mu$m, reflectivity $\rho = 17 \times 10^{-3}$, and self-coupling parameter $\sigma=0.98$.}
	\label{GammaTAU1}
\end{figure}

In \figref{ContourPLot}, we show the total emission rate normalized to the emission rate in a sole waveguide, presented as a function of the phase parameter $\Delta$ and reflectivity. We consider a point-dipole emitting at $\lambda_0=630 $ nm, a ring of radius length $l=300\pi\lambda_0=93.7$ $\mu$m, and $\sigma=0.98$.  As derived from \eqref{golden_rule_backscattering4}, the presence of backscattering dramatically impacts the emission rate, causing variations of several orders of magnitude depending on the relative positions of the dipole and the backscattering element. Notably, emission can be entirely suppressed when $\Delta = \pi/4 + m\pi$, where $m \in \mathbb{N}$. This significant change in emission arises from the onset of standing waves within the resonator, resulting from the coupling of clockwise and counterclockwise propagating modes. Consequently, emission is either strongly enhanced or suppressed based on whether the dipole is located at a maximum or a node of these standing waves. As expected, when $\rho\rightarrow 0$, one retrieves the result for the backscattering-free ring discussed in the previous section. 

Finally, in \figref{GammaTAU1}, we consider $\rho=17 \times 10^{-3}$ and plot the normalized dipole emission (black solid line) alongside the probability $P_{L(R)}=\Gamma_\mathrm{rb, L(R)}/\Gamma_\mathrm{rb}$ (red solid and dashed line, respectively) of the photon exiting the left (right) port of the structure. It is evident that, even for a small backscattering probability per round trip, where the overall dipole emission rate remains essentially independent of $\Delta$, the presence of backscattering still significantly impacts the spatial distribution of the emitted light. Therefore, such an effect must be carefully considered in the design and fabrication of single-photon sources based on dipole emission.

\section{\uppercase{Deterministic single-photon emission with an interferometric resonator}}\label{General}

In the previous section, we examined relatively simple yet instructive structures, demonstrating how our formalism provides a general and flexible framework for modelling spontaneous emission in integrated photonic systems. In particular, we showed that interference between clockwise and counterclockwise modes in ring resonators can have a pronounced effect on the spontaneous emission process. Building on these insights, we now take a step forward and consider a more complex photonic structure, designed to exploit and control such interference to achieve deterministic emission of a single photon into a well-defined mode. 

\begin{figure}[t!]
	\centering
 \includegraphics[width=\linewidth]{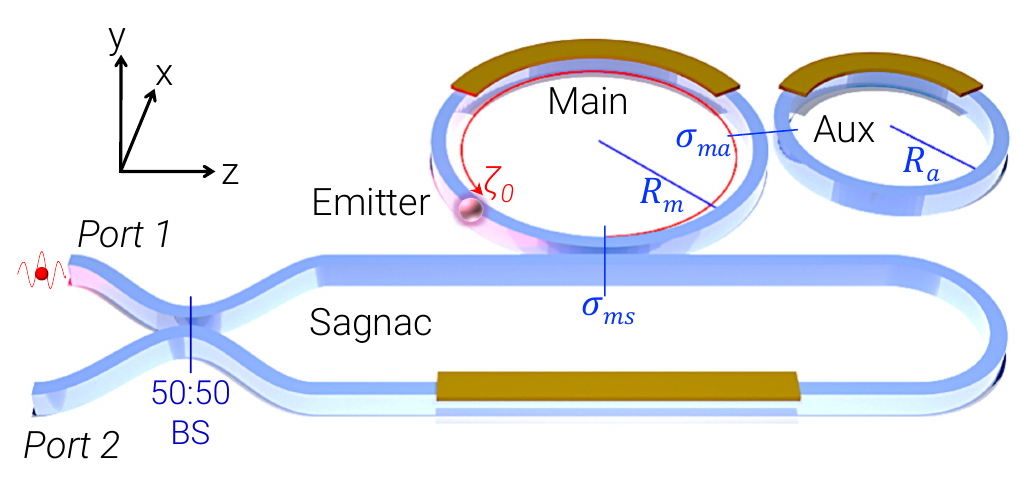}
	\caption{An illustration of the proposed device is shown. The system comprises a main ring resonator (Main) with radius $R_m$, which is coupled to a Sagnac interferometer (with a $50:50$ beam splitter) via a self-coupling parameter $\sigma_{ms}$, and also coupled to an auxiliary ring (Aux) of radius $R_a$ with coupling strength $\sigma_{ma}$. This configuration allows for tunable interference between the clockwise and counterclockwise propagating modes, thereby enabling control over the emission properties of the dipole (indicated by the red spot) embedded within the main ring.}
	\label{sagnac}
\end{figure}

As illustrated in \figref{sagnac}, the system under investigation consists of a photonic molecule formed by an auxiliary ring of radius $R_a$ and a main ring resonator of radius $R_m$ coupled to an integrated Sagnac interferometer of length $l_s$ with a $50:50$ beam splitter at the output ($\kappa_s=\sigma_s=1/\sqrt{2}$ see \appref{Appendix2}). The main resonator is assumed to host a single-photon emitter resonantly coupled to the main ring and operating at the wavelength $\lambda_0$. We assume that the resonant frequencies of the resonators, as well as the interference at the output of the Sagnac interferometer, can be controlled using phase shifters, such as heaters or electro-optic modulators, depending on the material platform employed \cite{tarasenko2007,liu2015,sinatkas2021}. 

As in the previous cases, the electromagnetic field is quantized in terms of the asymptotic-in fields of the structure, which correspond to the stationary solutions of Maxwell equations with light entering the structure either from port 1 or port 2. We emphasize once again that we are computing the probability for photons to \emph{exit} the structure. The choice of using asymptotic-in fields, rather than asymptotic-out fields, is made for mathematical convenience: the former are generally easier to compute. Following the general result of \eqref{fermi_golden}, we evaluate the asymptotic field in the main ring at the position of the dipole (see \appref{Appendix2}):
\begin{eqnarray}
   \mathbf{D}_{1,k_0}^\text{asy-in (main)}(\mathbf{r}_0)&=& f_{k_0}^{(1)}(\zeta_0) \frac{\mathbf{d}(\textbf{r}_{\perp,0},\zeta_0)}{\sqrt{2\pi}},  \\
   \mathbf{D}_{2,k_0}^\text{asy-in (main)}(\mathbf{r}_0)&=& f_{k_0}^{(2)}(\zeta_0) \frac{\mathbf{d}(\textbf{r}_{\perp,0},\zeta_0)}{\sqrt{2\pi}},
   \label{asymptotic3}
\end{eqnarray}
where $f_{k_0}^{(j)}(\zeta_0)$ quantifies the longitudinal field enhancement at the dipole position $(\textbf{r}_{\perp,0},\zeta_0)$ for the asymptotic field corresponding to the $j$-th output port \cite{Liscidini2012}.

From \eqref{fermigoldenrule2}, and using Eqs.~(\ref{Fielda_General_system}-\ref{Fielda_General_system2}), we can express the total emission rate as the sum of the emission rates from ports 1 and 2:
\begin{align}
    \Gamma_{\text{T}}
    = & \frac{\Gamma_{\text{wg}}}{2} \left( \left|f_{k_0}^{(1)}(\zeta_0)\right|^2  + \left|f_{k_0}^{(2)}(\zeta_0)\right|^2 \right) \nonumber \\
    = & \Gamma_{\text{wg}} \frac{\kappa_{ms}^2}{1 + \sigma_{ms}^2 + \sigma_{ms} \left[ C(\sigma_{ma},\delta_a) e^{i (\delta_a + \delta_m)} + \text{H.c.} \right]}\;,
\label{golden_rule_general_case}
\end{align}
where $\sigma_{ms}$ is the self-coupling coefficient between the Sagnac interferometer and the main ring, and $\sigma_{ma}$ is the coupling coefficient between the main and auxiliary resonators, with $\kappa_{ms(a)}=\sqrt{1-\sigma_{ms(a)}^2}$. The phase shifts per round trip in the main (auxiliary) rings is given by $\delta_{m(a)} = 2\pi k_0 R_{m(a)}$, and the function $C(\sigma_{ma},\delta_a) 
=(1 - \sigma_{ma} e^{-i \delta_a})/(1 - \sigma_{ma} e^{i \delta_a})$. It is worth noting that the total emission rate $\Gamma_{\text{T}}$ is independent of both the dipole position within the ring and the coupling point between the main ring and the Sagnac interferometer. Finally, in the limit $\sigma_{ma} \to 1$, \eqref{golden_rule_general_case} reduces to the single-ring result discussed in \eqref{fermigoldenrule3}.

The Sagnac interferometer enables control over the interference between emission into the clockwise and counterclockwise modes. As a result, the emitted photon is generally distributed unequally between port 1 and port 2. The probability of emission from the $j$-th port is given by
\begin{align} \nonumber
    P_j &= \frac{\Gamma_{j}}{\Gamma_{\text{T}}} \\
    &= \frac{1}{2} \left[ 1 - \frac{i^{2j+1}}{2}
    \left( C(\sigma_{ma},\delta_a)\, e^{i (\delta_a - \delta_s + \tilde{\delta}_0)} - \text{H.c.} \right) \right],
\label{golden_rule_general_case4}
\end{align}
where $\Gamma_j = \Gamma_{\text{wg}} \left|f_{k_0}^{(j)}(\zeta_0)\right|^2/2$ is the emission rate into the $j$-th port, $\delta_s = k_0 l_s$ is the phase shift associated with the Sagnac interferometer, and $\tilde{\delta}_0 = 2k_0(\zeta_0 + l_1)$ is a phase term that depends on the longitudinal dipole position $\zeta_0$ and the distance $l_1$ between the Sagnac beam splitter and the coupling point of the main ring to the Sagnac interferometer (see \figref{Figure2} in Appendix~\ref{Appendix2}). From \eqref{golden_rule_general_case4} it is clear that the output distribution can be engineered by tuning structural parameters and phase shifts, enabling control over the emission intensity and directionality. 
\begin{figure}[ht!]
	\centering
 \includegraphics[width=\linewidth]{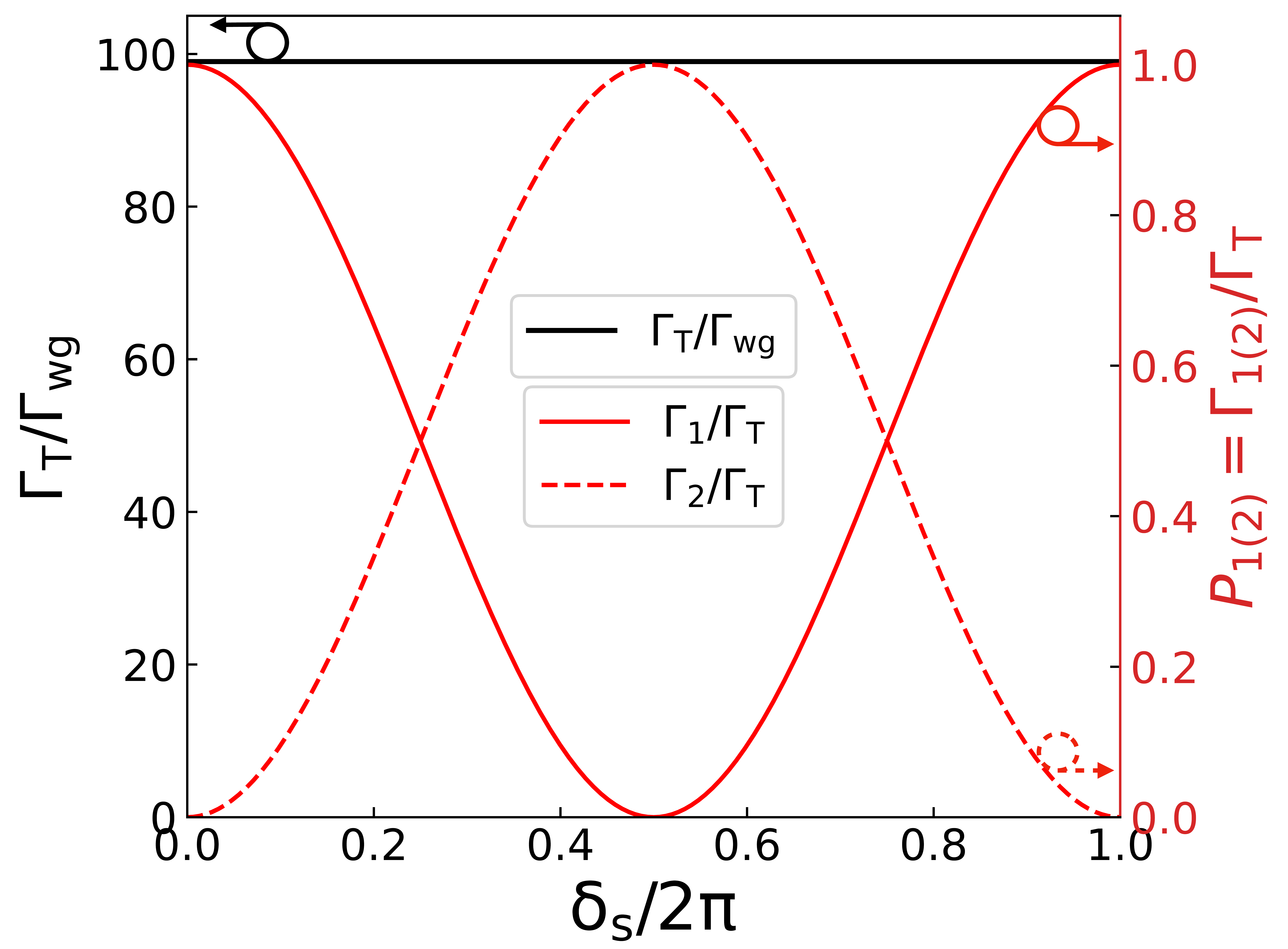}
	\caption{The normalized dipole emission rate (black solid line), $\Gamma_{{\text{T}}}/\Gamma_{\text{wg}}$, and the normalized output probability, $P_j$ (red solid and dashed line, respectively), are shown as functions of the Sagnac interferometer phase $\delta_s$. In particular, when \( \delta_s =\{ 0,2\pi \} \), the photon always exits from port ~1, while for \( \delta_s = \pi \), it exits from port ~2. By continuously adjusting \( \delta_s \), it is possible to generate any coherent superposition of the photon exiting the two ports. The parameters used are: $\lambda_0=630\;\text{nm}$ , $l=300\pi\lambda_0=93.7$ $\mu$m,, $R_m = 158\lambda_0$, $R_a = R_m/2$, which yield $\delta_m = \delta_a = 2\pi$ and $\tilde{\delta}_0 = \pi/2$. The self-coupling parameters are set as $\sigma_1 = 1/\sqrt{2}$, $\sigma_{ms} = 0.98$, and $\sigma_{ma} = 0.7$.}
	\label{Figure4}
\end{figure}
\begin{figure}[ht!]
	\centering
 \includegraphics[width=\linewidth]{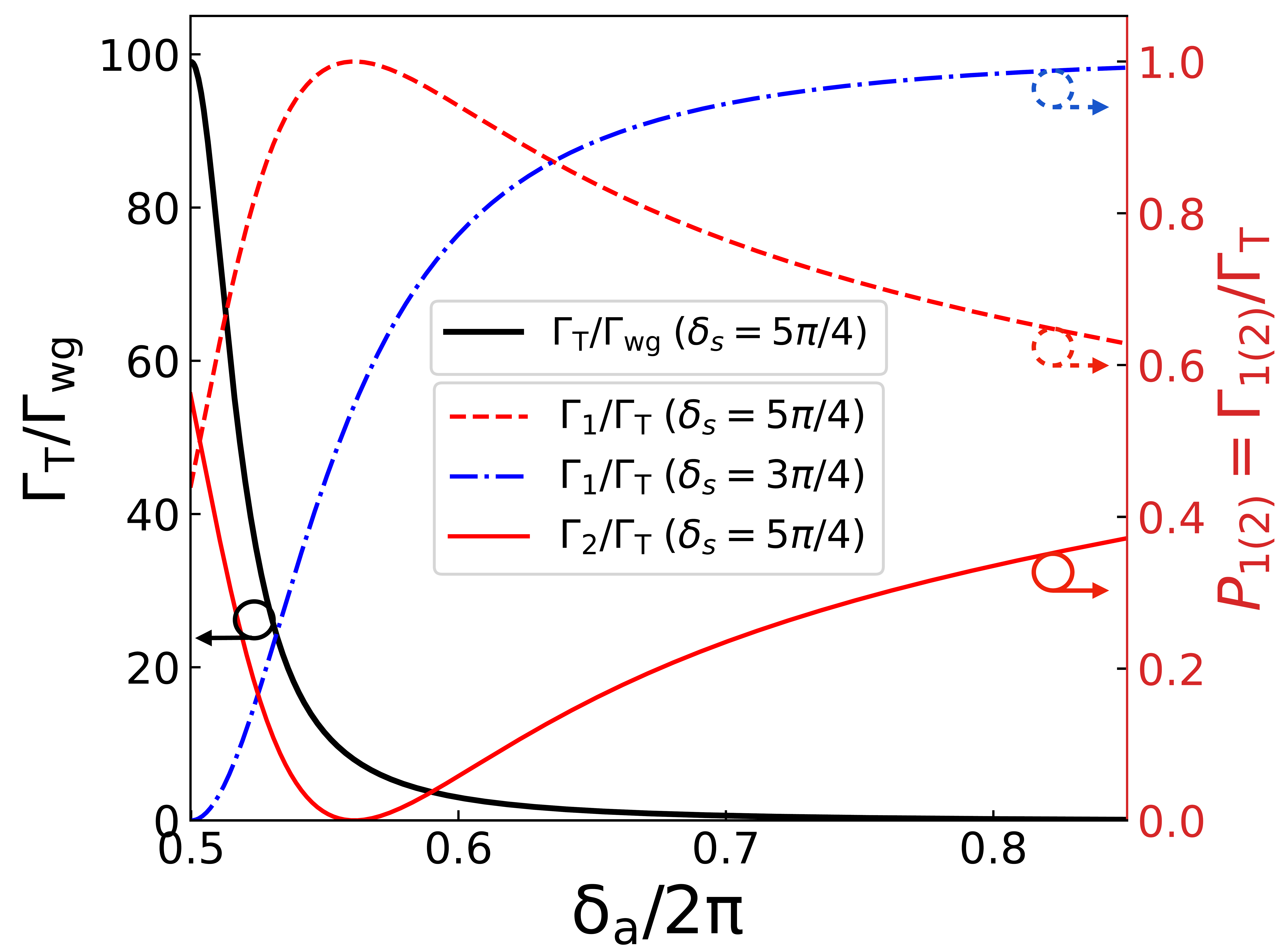}
	\caption{Normalized dipole emission rate $\Gamma_{{\text{T}}}/\Gamma_{\text{wg}}$ and normalized output probability $P_j$ as function of the auxiliary ring phase $\delta_a$. By fixing $\delta_s= 5\pi /4$, the probability of having a superposition state of photons emitted is close to one (red dashed and continuous curve) and can be controlled by varying $\delta_a$, suppressing in the meantime the dipole emission (black continuous curve). For $\delta_s= 3\pi/4$ the dipole emission is still suppressed by adjusting $\delta_a$, while the photon emission probability from port 1 is enhanced until it reaches the value of one (blue dot-dashed curve). Other parameters used are: $\lambda_0=630\;\text{nm}$ , $l=300\pi\lambda_0=93.7$ $\mu$m,, $R_m = 158\lambda_0$, $R_a = R_m/2$, so that, $\delta_m=2\pi$, $\tilde{\delta}_0=\pi/2$. Self-coupling parameters are taken as $\sigma_1=1/\sqrt{2}$, $\sigma_{ms}=0.98$, $\sigma_{ma}=0.7$.}
	\label{Figure5}
\end{figure}
First, we consider the dependence of the normalized dipole emission and the output photon probabilities on the Sagnac phase shift \( \delta_s \). In particular, \figref{Figure4} shows the normalized emission rate into the main resonator, \( \Gamma_{\text{T}}/\Gamma_{\text{wg}} \) (black solid line), along with the probabilities of photon emission from port 1 and port 2, \( \Gamma_{1}/\Gamma_{\text{wg}} \) and \( \Gamma_{2}/\Gamma_{\text{wg}} \) (red solid and dashed lines, respectively), as functions of \( \delta_s \).

Assuming a main ring radius of \( R_m = 100~\mu\text{m} \) and an intrinsic quality factor \( Q = 10^5 \) (limited only by propagation losses, which are included in the calculation of the asymptotic fields \cite{banic22}), the structure achieves an extraction efficiency of approximately 99\%, as indicated by the nearly constant value of \( \Gamma_{\text{T}}/\Gamma_{\text{wg}} \). This high efficiency, maintained across all values of \( \delta_s \), reflects the strong enhancement of spontaneous emission — about two orders of magnitude greater than that of a simple waveguide — ensuring that the dipole primarily emits into the guided mode of the main ring.

The Sagnac phase shift \( \delta_s \) plays a crucial role in determining the photon's emission direction. When \( \delta_s =\{ 0,2\pi \} \), the photon always exits from port~1, whereas for \( \delta_s = \pi \), it exits from port~2 (\figref{Figure4}). By continuously tuning \( \delta_s \), it is possible to generate any coherent superposition of the photon exiting from the two ports. This tunability allows full control over the photon's path degree of freedom, without requiring knowledge of the dipole's precise longitudinal position within the ring.

Finally, we discuss how this structure can be used to control both the dipole emission rate and the output probability at each port depending on the auxiliary ring phase shift \( \delta_a \). As illustrated in \figref{Figure5}, by tuning \( \delta_a \), one can selectively enhance or suppress the dipole's emission rate. This effect stems from the strong coupling between the main and auxiliary resonators, which allows shifting the resonance frequency of the main ring and thereby controlling the dipole-mode interaction strength. When the coupling is strong and induces a full splitting of the main resonance, the dipole becomes off-resonant with both hybrid modes, leading to a suppression of spontaneous emission. Conversely, when the auxiliary ring is far-detuned and effectively uncoupled, the dipole remains resonant with the main ring, resulting in efficient photon emission.
The presence of the auxiliary resonator influences both the emission rate and the quantum state of the emitted photons, which - in the path degree of freedom - is determined by the probability of exiting from one of the two ports of the structure.
Yet, by adjusting both the values of $\delta_a$ and $\delta_s$, one can achieve the simultaneous control of the emission rate and the emission path. As an example, we show the trend for  $\delta_s = 5\pi/4$ and $\delta_s = 3\pi/4$. 

\section{Conclusions}\label{conclu}
In this work, we have introduced a general and versatile framework for modeling dipole emission in arbitrary integrated photonic structures, grounded in the quantization of the electromagnetic field using asymptotic in/out mode formalism. By building directly on the stationary solutions of Maxwell’s equations, this method circumvents common approximations—such as assuming Lorentzian lineshapes or idealized point-like system-bath couplings—that often limit the accuracy of conventional approaches. Its applicability extends across a wide range of photonic geometries and material platforms, including dispersive, lossy, and non-Hermitian systems. As a result, the framework offers a powerful and reliable tool for analyzing spontaneous emission in complex nanophotonic environments, enabling precise insights into light–matter interaction in integrated quantum devices. This paves the way for more predictive and design-oriented modeling in emerging quantum photonic technologies.

In \secref{model}, we applied our approach to increasingly complex structures, from a single-mode waveguide to a ring resonator, and then to a ring with backscattering. In each case, our formalism not only reproduces known results, such as the Purcell enhancement in high-finesse cavities, but it also provides direct physical insight into the mechanisms that govern emission control, including the roles of transverse field confinement, group velocity, and the longitudinal resonant buildup in a resonant structure. Importantly, effects such as emission suppression due to mode splitting or standing wave formation emerge naturally within our approach, without the need for additional phenomenological models.

In \secref{General}, we demonstrated the potential of this framework through the design and analysis of a tunable single-photon source based on a main ring resonator coupled both to an auxiliary ring and a Sagnac interferometer. We showed that the emission rate can be dynamically controlled by exploiting the hybridization of the ring modes, while the output port of the emitted photon can be independently selected by tuning the Sagnac phase. This simultaneous control over both the emission rate and the spatial mode occupation illustrates the power of our method as a tool for the design of quantum light sources.

Beyond the specific cases presented here, the asymptotic-field formalism applies to arbitrary integrated geometries and configurations, including multi-mode waveguides, multi-port structures, and multiple emitters. The ability to identify emission pathways and rates channel by channel makes this approach particularly appealing for applications in quantum photonics, where precise control over single-photon generation, routing, and interference is essential.

Looking ahead, we envision this method becoming a valuable tool for the modeling and design of integrated quantum photonic devices.  It offers a powerful framework not only for interpreting the underlying physics of light–matter interactions at the nanoscale but also for guiding the development of robust and scalable quantum technologies.  This could be the case for applications in  T-centers architecture, which is optimized for overall entanglement distribution \cite{Simmons2024}, and by enabling precise control over emission dynamics and interference effects, this approach holds promise for advancing applications in quantum information processing \cite{flamini2018photonic}, communication \cite{bennett2014quantum}, and sensing \cite{zhang2021distributed}.

\acknowledgments
V. M and M.L. acknowledge PNRR MUR project “National Quantum Science and Technology Institute” – NQSTI (Grant No. PE0000023).

\appendix
\section{From Fermi's golden rule to dipole emission in terms of asymptotic fields} \label{appendixa}

In this appendix, we derive a general expression for the dipole emission rate by applying the asymptotic field formalism within the framework of the Fermi Golden Rule. This approach allows us to systematically account for the contributions of both forward- and backward-propagating modes, providing a comprehensive description of the dipole’s radiative behavior in structured photonic environments. 

Since we are interested in spontaneous emission, we consider as initial state $|i_{N,k}\rangle=|n_{N,k}\rangle \otimes|e\rangle$, where $|n_{N,k}\rangle$ represents a photon number state with $n$ excitations in the $N$-th channel, and $|e \rangle$ is the excited state of the dipole. The total energy of this initial state is $E_i=\hbar \omega_e + \hbar n \omega_k$, where $\hbar \omega_e$ is the energy of the excited state of the dipole, and $\hbar \omega_k$ is the energy of each photon in mode $k$. Under the effect of the interaction the final state is given by $|f_{N,k}\rangle=|n_{N,k}+1\rangle \otimes|g\rangle$, where the dipole has relaxed to its ground state $|g\rangle$ with energy $\hbar \omega_g$, and one additional photon has been created in the ring resonator. Therefore, the total energy of this final state is $E_f = \hbar \omega_g + \hbar (n+1) \omega_k $.  

By using  \eqref{interactionI} and \eqref{interactionII}, the total emission rate in \eqref{fermigoldenrule} can be expressed explicitly in terms of the asymptotic field components and their coupling to the dipole as follows
\begin{widetext}
\begin{align}
    \Gamma =& \frac{2\pi}{\hbar}  \sum_N \int \textrm{d}k \biggr{|} \langle f_{N,k}|\hat{\mathbf{p}} \cdot \hat{\mathbf{E}}(\textbf{r}_0)|i_{N,k}\rangle \biggr{|}^2 \delta(\hbar  \omega_k- \hbar  \omega_0) \; \nonumber \\
    =& \frac{2\pi}{\varepsilon_0^2 n^4(\mathbf{r_0}) \hbar} \sum_{N} \int \textrm{d}k \biggr{|}   \sum_{N'} \int \textrm{d}k' \sqrt{\frac{\hbar\omega_{N',k'}}{2}} \langle f_{N,k}| \biggr{(} \hat{\sigma}^+ \hat{a}_{N',k'} \biggr{[} \mathbf{p} \cdot \mathbf{D}^\text{asy-in}_{N',k'}(\mathbf{r}_0)\biggr{]} + \text{H.c.}\biggr{)}|i_{N,k}\rangle \biggr{|}^2 \delta(\hbar  \omega_{N,k}- \hbar  \omega_0) \;,
    \label{fermigoldenrule1}
\end{align}
\end{widetext}
where we defined $\omega_0= \omega_e - \omega_g$ and applied the rotating wave approximation to the dipole interaction. For each channel, considering a sufficiently narrow dipole emission, we introduce the usual dispersion relation expanded in Taylor series up to the first order, around the given transition frequency $\omega_0$, which is chosen depending on the dipole spectral emission \cite{Sipe2004,quesada2022}, 
\begin{equation}
k(\omega)=k_{0,N}+\frac{1}{{v}_{g_{N}}}(\omega-\omega_0)+... \, 
\label{Taylor}
\end{equation}
Here, ${v}_{g_{N}}$ is the group velocity of the electric field propagating through the $N$-th channel and $k_{0,N}=\omega_0 n(\mathbf{r_0})/c$ with  $n(\mathbf{r_0})$ being the refraction index. The action of the initial and final ket on the interaction terms gives $\langle f_{N,k}|\hat{\sigma}^+ \hat{a}_{N',k'}|i_{N,k}\rangle=\delta_{N,{N'}}\delta(k'-k) \sqrt{n_{{N'},k'}+1}$, so that, assuming that the two vectors $\mathbf{p}$ and $\mathbf{D}^\text{asy-in}_{N,k}(\mathbf{r}_0)$ are aligned, we can estimate the integral in \eqref{fermigoldenrule1} arriving to
\begin{eqnarray}
\Gamma = \frac{ \pi \omega_0 }{\varepsilon_0^2 n^4(\mathbf{r_0}) \hbar} \sum_{N} \frac{n_{{N},k_{0}} + 1}{{v}_{g_{N}}}\left|\mathbf{p} \cdot \left[\mathbf{D}^\text{asy-in}_{N,k_{0,N}}(\mathbf{r}_0)\right]^*\right|^2\;.
\label{fermigoldenrule4}
\end{eqnarray}
Here, $\left[\mathbf{D}^\text{asy-in}_{N,k}(\mathbf{r}_0)\right]^*$ is the conjugate of asymptotic-in field (for the $N$-th channel) which happens to be the asymptotic-out field \cite{Liscidini2012}, namely, $\left[\mathbf{D}^\text{asy-in}_{N,k}(\mathbf{r}_0)\right]^*=\mathbf{D}^\text{asy-out}_{N,k}(\mathbf{r}_0)$.

The \eqref{fermigoldenrule2} is derived from \eqref{fermigoldenrule4} by assuming that there are no photons in any of the resonator channels in the initial state, i.e., $n_{{N},k_{0}}=0$, and it gives the probability (per unit time) that a single photon is emitted by a dipole interacting with an electromagnetic field confined along one dimension of an isotropic optical material.

\section{Waveguide coupled to a ring resonator: an asymptotic-in/out fields approach}\label{Appendix1}
\begin{figure}[ht!]
	\centering
\includegraphics[width=1\linewidth]{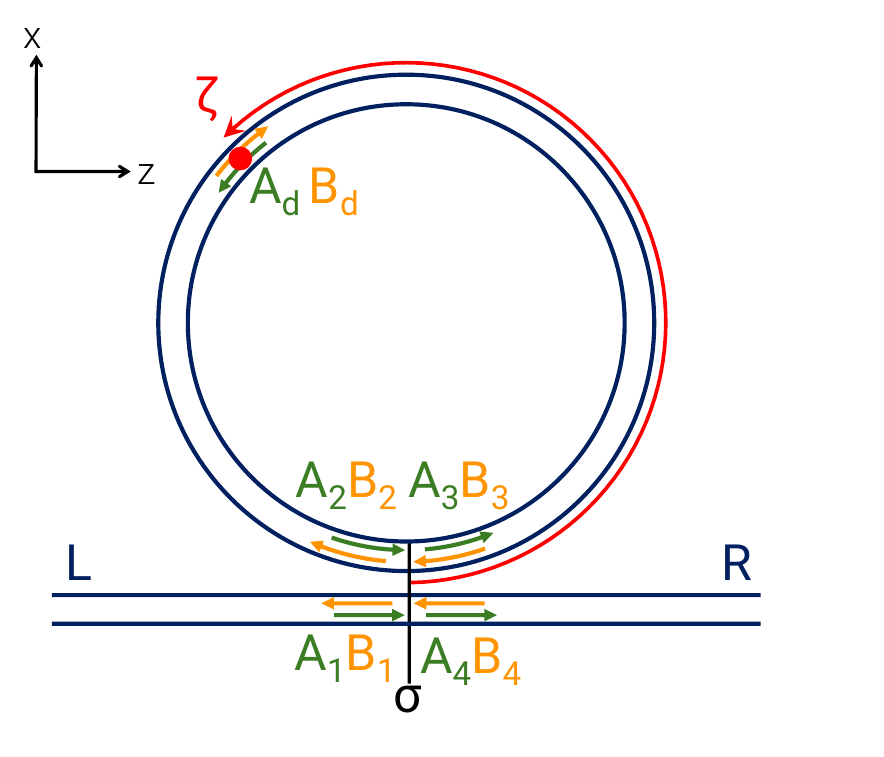}
	\caption{Schematic representation of a point-coupled ring resonator (of radius $R$) with a waveguide. The coupling is modeled by a simple scattering matrix, with self-, $\sigma$ and cross-coupling, $\kappa$, coefficients, respectively. $\{A_i,B_i\}$, with $i=1,..4$,  are the fields before and after the coupling point, while $\{A_d,B_d\}$ are the clockwise and counterclockwise field interacting with the dipole (the red dot at the position $\zeta$) inside the ring.}
	\label{only_ring_resonator}
\end{figure}

In order to derive the correct generic form for the scattering part of the asymptotic field $\mathbf{D}^\text{asy-in}_{N,k}(\textbf{r})$, we have to solve a linear equations system following the scheme in  \figref{only_ring_resonator}. With the notation $\{A_i,B_i\}$ ($i=1,..4$) we refer to the incoming and outgoing fields at coupling point, while with $\{A_d,B_d\}$ we indicate the counterclockwise and clockwise %clockwise and counterclockwise
field interacting with the dipole (the red dot at the position $\zeta$) inside the ring. Here, we treat the coupling between the channels and the ring in the usual way, with self-, $\sigma$, and cross-, $\kappa$, coupling coefficients satisfying $\sigma^2 +\kappa^2= 1$, such that the coupler is described by the set of equations
\begin{equation}\label{linear_system_one_ring_resonator}
\begin{cases}
            \sigma A_1 + i \kappa A_2 - A_4=0 \\
             i \kappa A_1 +  \sigma A_2 - A_3=0\\
            \sigma B_1 - i \kappa B_2 - B_4=0 \\
             -i \kappa B_1 +  \sigma B_2 - B_3=0. 
        \end{cases}
\end{equation}
To this, one has to add the shift equations which are given by
\begin{equation}\label{linear_system_one_ring_resonator2}
\begin{cases}
            A_2-A_3e^{i \delta}=0  \\
             B_2-B_3e^{-i \delta}=0 ,
        \end{cases}
\end{equation}
where $\delta= k l $, being $l=2\pi R $ with $R$ the ring radius.

One can solve the  system equations \eqref{linear_system_one_ring_resonator} together with  \eqref{linear_system_one_ring_resonator2}, under two possible initial conditions, which represent the counterclockwise and clockwise asymptotic-in fields in the ring. If light enters from the left (L) port, we have $\{A_1=1,B_4=0\}$,  and thus $B_2 = 0$ and
\begin{equation}
         A_3 = \frac{i \kappa}{1 - \sigma e^{i \delta}} \;.
\end{equation}
If light enters from the right (R) port, we have $\{A_1=0,B_4=1\}$,  and thus $A_3 = 0$ and
\begin{equation}
         B_2 = \frac{i \kappa}{1 - \sigma e^{i \delta}} \;.
\end{equation}
We note that the two asymptotic fields have the same expressions due to the symmetry of this structure.

In particular, we are interested in $A_d = A_3 e^{i \tilde{\delta}}$ and $B_d = B_2 e^{i(k l - \tilde{\delta})}$ 
with $\tilde{\delta}=k\zeta$ and $\zeta$ the dipole coordinate in the counterclockwise direction along the ring circumference. Indeed, $B_d$ and $A_d$ are the scattering parts of the asymptotic clockwise and counterclockwise fields at the dipole position, respectively. Note that, at the resonant condition $k l = 2m\pi$, with $m$ integer, one can write $B_d = B_2 e^{-i \tilde{\delta}}$. Therefore, we have that the two asymptotic fields at the dipole position are
\begin{align}
    {\mathbf{D}_{L,k}^\text{asy-in (ring)}
}(\mathbf{r})=&  \frac{i \kappa  e^{i \tilde{\delta}}  }{1-\sigma e^{i \delta}}~\frac{\mathbf{d}(\textbf{r}_{\perp},\zeta)}{\sqrt{2\pi}} \;, \\
    {\mathbf{D}_{R,k}^\text{asy-in (ring)}}(\mathbf{r})=&  \frac{i \kappa e^{i(k l- \tilde{\delta})}}{1-\sigma e^{i \delta}}~\frac{\mathbf{d}(\textbf{r}_{\perp},\zeta)}{\sqrt{2\pi}}\;, 
    \label{asymptotic1}
\end{align}
where $\mathbf{d}(\textbf{r}_{\perp},\zeta)$ is the transverse field profile at the dipole position, which has radial coordinate $\textbf{r}_{\perp}$.

\section{Waveguide coupled to a ring resonator with backscattering: an asymptotic-in/out fields approach}
\label{Appendix_backscattering2}

\begin{figure}[ht!]
	\centering
 \includegraphics[width=1\linewidth]{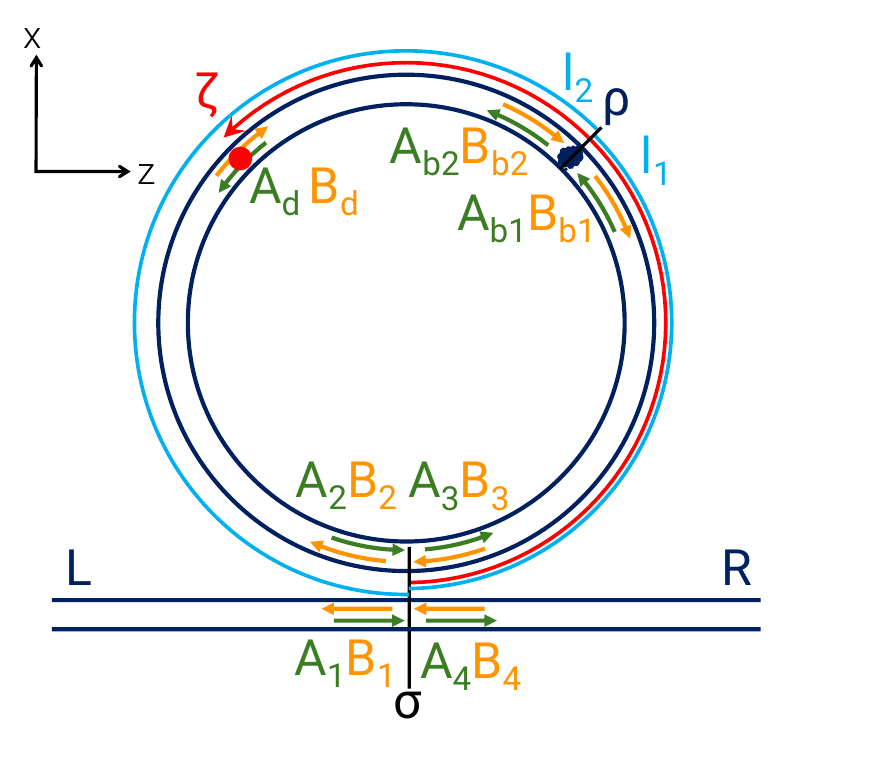}
	\caption{Schematic representation of a ring resonator (of radius $R$) point-coupled to a waveguide and containing a scatterer (blue spot inside the ring) with $\rho$ as the backscattering probability amplitude per round trip. The coupling is modeled by a simple scattering matrix, with self-, $\sigma$ and cross-coupling, $\kappa$, coefficients.  $\{A_i,B_i\}$, with $i=1,..4$,  are the fields before and after the coupling point, $\{A_{bi},B_{bi}\}$, with $i=1,2$, are the fields before and after the backscattering point (the blue dot inside the ring). Finally, $\{A_d,B_d\}$ are the fields interacting with the dipole (the red dot at the position $\zeta$).}
	\label{backscattering_ring_resonator}
\end{figure}

We now extend the derivation presented in \appref{Appendix1} to account for the presence of a defect—represented by the blue dot in \figref{backscattering_ring_resonator}—within the ring resonator. This defect introduces backscattering, coupling the clockwise and counterclockwise propagating fields.
With $\{A_{b1},B_{b1}\}$ we refer to the counterclockwise and clockwise fields before the lumped scattering point, whereas $\{A_{b2},B_{b2}\}$ are the counterclockwise and clockwise fields after the lumped scattering point, where we define \textit{before} and \textit{after} according to the counterclockwise direction. For the other fields in play, we use the same notation as before.

In addition to the four equations for the coupler \eqref{linear_system_one_ring_resonator}, we have now two more equations coupling the clockwise and counterclockwise fields at the scatterer position 
\begin{equation}
 \label{linear_system_backscattering_ring_resonator3}
    \begin{cases}
        \tau A_{b1} + i \rho B_{b2} - A_{b2} =0 \\
        i\rho A_{b1} + \tau B_{b2} - B_{b1} =0 \\ 
    \end{cases} 
\end{equation}
where $\rho$ is the backscattering probability amplitude per round trip, and $\tau = \sqrt{1-\rho^2}$.
The new shift equations are given by
\begin{equation}
    \begin{cases}
        A_{b1} - A_3 e^{i \delta_1}=0  \\
        B_{b1} - B_3 e^{-i \delta_1}=0  \\
        A_2 - A_{b2} e^{i \delta_2 }=0  \\
        B_2 - B_{b2} e^{-i \delta_2}=0  \\
    \end{cases} 
\end{equation}
where $\delta_{(1,2)}=kl_{(1,2)}$ (with $l=l_1+l_2=2\pi R$) are the phase shifts along the arc spanning from the coupling point to the lumped backscattering in the clockwise and counterclockwise direction (see \figref{backscattering_ring_resonator}).

Differently from what we have seen in the previous simple case, due to the presence of the backscattering point, now the asymptotic fields felt by the dipole are always a superposition of a clockwise and counterclockwise electromagnetic fields $f_{k}^{(L,R)}(\zeta)=A_d^{(L,R)} + B_d^{(L,R)}=A_{b2} e^{i (\tilde{\delta}-\delta_1)}+ B_2 e^{i(\delta-\tilde{\delta})}$, regardless of whether the photon is injected from the left  $\{A_1=1, B_4=0 \} $ or the right $\{A_1=0, B_4=1 \}$. Here, $\tilde{\delta}=k \zeta$ with $\zeta$ being the angular position of the dipole, while  $\delta=\delta_1 + \delta_2 =kl$. Therefore, by solving \eqref{linear_system_backscattering_ring_resonator3}-\textcolor{blue}{(C2)} under the first condition $\{A_1=1, B_4=0\}$, we obtain the first solution  
\begin{align}
f_{k}^{(L)}(\zeta) =  \frac{i \kappa (\tau - \sigma e^{i \delta}) e^{i \tilde{\delta}} - \rho \kappa \sigma  e^{2 i \delta_1} e^{ i (\delta-\tilde{\delta})}}{1- 2 \tau \sigma e^{i \delta} + \sigma^2 e^{2 i \delta}}\;,  
\label{Fielda_Back_system}
\end{align}
whereas under the second initial condition $\{A_1=0, B_4=1\}$,  we obtain 
\begin{align}
f_{k}^{(R)}(\zeta) =  \frac{i \kappa (1- \tau \sigma e^{i \delta}) e^{i (\delta - \tilde{\delta})} - \rho \kappa  e^{ i \delta_2} e^{ i (\tilde{\delta} - \delta_1)}}{1- 2 \tau \sigma e^{i \delta} + \sigma^2 e^{2 i \delta}}\;,  
\label{Fieldb_back_system}
\end{align}

Finally, the asymptotic-in fields inside the main ring are given by composing \eqref{Fielda_Back_system} and \eqref{Fieldb_back_system}, so that 
\begin{align}
    {\mathbf{D}_{(L,R),k}^\text{Ring}}(\textbf{r}_{\perp},\zeta)=&  (A_d^{(L,R)} + B_d^{(L,R)})~\frac{\mathbf{d}_{(1,2),k}(\textbf{r}_{\perp},\zeta)}{\sqrt{2\pi}} \nonumber\\ 
    = & f_{k}^{(L,R)}(\zeta) \frac{\mathbf{d}_{(L,R),k}(\textbf{r}_{\perp},\zeta)}{\sqrt{2\pi}} \;. 
    \label{back_General}
\end{align}
Note that, we use  $\textrm{d}\textbf{r}=\textrm{d}\textbf{r}_{\perp}\textrm{d}\zeta$, where $\textbf{r}_{\perp}$ indicates the coordinates of the dipole position in the plane normal to the ring resonator, while with $\zeta$ we refer to the coordinate position in the counterclockwise direction along the ring circumference.

\begin{figure*}[t!]
	\centering
 \includegraphics[width=0.9\linewidth]{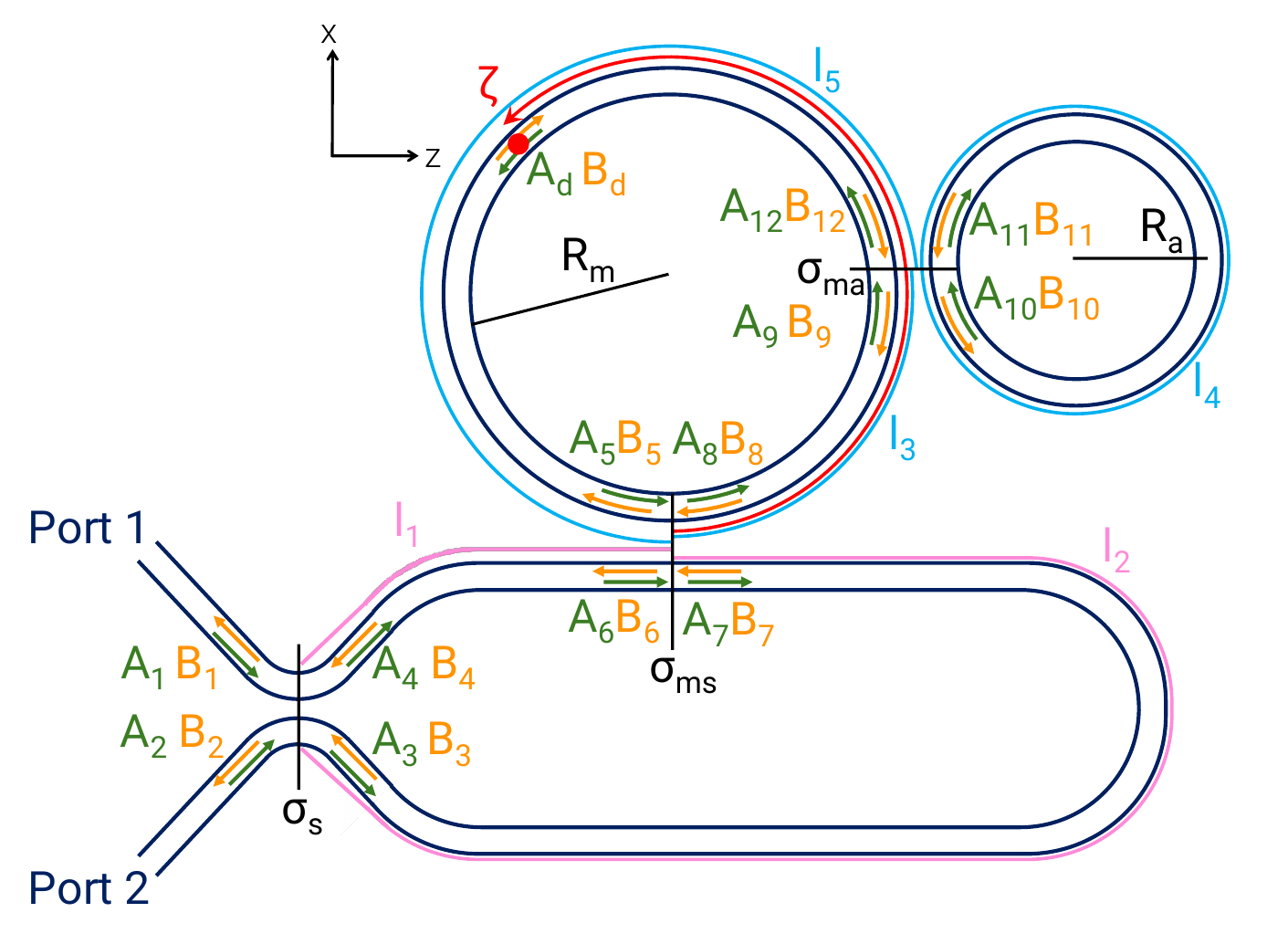}
	\caption{Schematic representation of a multi point-coupled ring resonator (radius $R_m$) with an auxiliary ring resonator (radius $R_a$), and an integrated Sagnac interferometer. The coupling in each point (inside the blue box) is modelled by a scattering matrix, with the self-, $\sigma_m (k)$ and cross-coupling, $\kappa_m(k)$, coefficients  ($m=1,2,3$), respectively. The $\{A_i,B_i\}$, with $i=1,...,12$,  are the fields before and after the interaction with the each coupling point inside the blue boxes, while $\delta_i=k l_i$ ($i=1,..,5$), being $l_i$ light path lengths, take into account the phase shift along the integrated system. The scattering part of the asymptotic clockwise and counterclockwise fields at the dipole position (the red dot inside the main ring) is the sum of the fields $\{A^{(1,2)}_d $, $B^{(1,2)}_d\}$ for each initial condition, namely, light incoming from channel 1 or channel 2.}
	\label{Figure2}
\end{figure*}

\section{Interferometric waveguide coupled to a ring resonator: an asymptotic-in/out fields approach}\label{Appendix2}

Here, we extend the method employed above to a more general geometry sketched in  \figref{Figure2}. We consider an integrated Sagnac
interferometer coupled to a main ring resonator with radius $R_m$, which is itself coupled to an auxiliary ring resonator with radius $R_a$. The two-level system is placed in the main ring resonator (the red dot). In order to calculate the dipole emission rate $\Gamma$ in this general structure, we have to calculate the scattering part of the asymptotic-in/out clockwise and counterclockwise fields that propagate through the main ring at the dipole position. 

Due to the presence of multiple point couplings, the scattered component of the asymptotic clockwise and counterclockwise fields at the dipole position—also in this case—is given by the sum of the fields $\{A^{(1,2)}_d $, $B^{(1,2)}_d\}$ corresponding to each initial condition, namely, light incoming from channel 1 or channel 2. In order to calculate the scattering matrix, we need to solve the system of algebraic equations for the incoming and outgoing fields at the coupling points $\{A_i,B_i\}$ ($i=1,...,12$), which are given by any possible path indicate in \figref{Figure2}. All the fields are connected each other by the self-, $\sigma_i$, and cross-, $\kappa_i$ coupling coefficients ($i=\{s,ms,ma\}$) which are supposed $k$-independent. Furthermore, each output field from a point coupling is connected to the input field of the next point coupling by a phase factor $e^{i \delta_j}$ ($j=1,..,5$).  

Therefore, the system equations for the sagnac-sagnac point coupling are given by  
\begin{equation}\label{linear_system_dissipation_model1}
 \begin{cases}
            \sigma_s A_1 + i \kappa_s A_2 - A_4=0 \\
             i \kappa_s A_1 +  \sigma_s A_2 - A_3=0\\
            \sigma_s B_1 - i \kappa_s B_2 - B_4=0 \\
             -i \kappa_s B_1 +  \sigma_s B_2 - B_3=0\;,  
        \end{cases} 
\end{equation}
while the one for the sagnac-main-ring point coupling is 
\begin{equation}\label{linear_system_dissipation_model2}
 \begin{cases}
            \sigma_{ms} A_5 + i \kappa_{ms} A_6 - A_8=0 \\
             i \kappa_{ms} A_5 +  \sigma_{ms} A_6 - A_7=0\\
            \sigma_{ms} B_5 - i \kappa_{ms} B_6 - B_8=0 \\
             -i \kappa_{ms} B_5 +  \sigma_{ms} B_6 - B_7=0\;,  
        \end{cases}
\end{equation}
and, finally, that one for the main-ring-aux-ring point coupling is 
\begin{equation}\label{linear_system_dissipation_model3}
        \begin{cases}
            \sigma_{ma} A_9 + i \kappa_{ma} A_{10} - A_{12}=0 \\
             i \kappa_{ma} A_9 +  \sigma_{ma} A_{10} - A_{11}=0\\
            \sigma_{ma} B_9 - i \kappa_{ma} B_{10} - B_{12}=0 \\
             -i \kappa_{ma} B_9 +  \sigma_{ma} B_{10} - B_{11}=0\;. 
        \end{cases}
\end{equation}
Moreover, one needs to take into account each phase shift $\delta_j=k l_j$ ($i=1,..,5$) being $l_j$ light path lengths (see \figref{Figure2}). Thus, we end up with two other systems of equations involving the shifted fields
\begin{equation}\label{linear_system_dissipation_model4}
\begin{cases}
            A_6-A_4 e^{i \delta_1}=0 \\
            B_7-A_3 e^{i \delta_2}=0 \\
            A_9-A_8 e^{i \delta_3} =0 \\
            A_{10}-A_{11} e^{i \delta_4}=0 \\
            A_5-A_{12} e^{i \delta_5}=0\;,
        \end{cases}
\end{equation}
and
\begin{equation}\label{linear_system_dissipation_model5}
      \begin{cases}
            B_4-B_6 e^{i \delta_1}=0  \\
             B_3-A_7 e^{i \delta_2}=0 \\
            B_8-B_9 e^{i \delta_3}=0  \\
            B_{11}-B_{10} e^{i \delta_4}=0 \\
            B_{12}-B_5 e^{i \delta_5}=0\;.
        \end{cases}
\end{equation}

The scattering part of the asymptotic field interacting with the dipole is given by $f_{k}^{(1,2)}(\zeta)=A^{(1,2)}_d + B^{(1,2)}_d = A_{12}e^{i (\tilde{\delta} - \delta_3)}  + B_5 e^{i(2 k \pi R_m -\tilde{\delta})} $, in both scenarios:  when a photon is injected into channel 1 $\{A_1=1, A_2=0\}$, and when a photon arrives from channel 2 $\{A_1=0, A_2=1\}$. Here, one has  $\tilde{\delta}=k \zeta$ with $\zeta$ being the angular position of the dipole, so that, in resonant condition $k=1/R_m$ one gets $e^{i(2 k \pi R_m -\tilde{\delta})}=e^{-i\tilde{\delta}}$.
Therefore, by solving \eqref{linear_system_dissipation_model1}-\textcolor{blue}{(D5)} under the first condition $\{A_1=1, A_2=0\}$, we obtain the first solution  
\begin{align}
f_{k}^{(1)}(\zeta) =& - \frac{i\sigma_s \kappa_{ms}  e^{i( \delta_1 +  \delta_3 +  \delta_4)} (1 - \sigma_{ma} e^{-i \delta_4})e^{i (\tilde{\delta} -\delta_3)}}{1- \sigma_{ma} e^{i \delta_4} + \sigma_{ms} e^{i (\delta_3 + \delta_4 + \delta_5)} (1- \sigma_{ma} e^{-i \delta_4})}  \nonumber \\
&  -  \frac{ \kappa_s \kappa_{ms}  e^{i \delta_2 } (1 - \sigma_{ma} e^{ i  \delta_4}) e^{ - i \tilde{\delta}}}{1- \sigma_{ma} e^{ i \delta_4} + \sigma_{ms} e^{i (\delta_3 + \delta_4 + \delta_5)} (1 - \sigma_{ma} e^{ - i  \delta_4})}\;,  
\label{Fielda_General_system}
\end{align}
whereas under the second initial condition $\{A_1=0, A_2=1\}$,  we obtain 
\begin{align}
f_{k}^{(2)}(\zeta)=&  \frac{ \kappa_s \kappa_{ms} e^{i ( \delta_1 +  \delta_3 +  \delta_4)} (1 - \sigma_{ma} e^{- i  \delta_4}) e^{i (\tilde{\delta} - \delta_3)} }{1- \sigma_{ma} e^{i \delta_4} + \sigma_{ms} e^{i (\delta_3 + \delta_4 + \delta_5)} (1- \sigma_{ma} e^{-i \delta_4})} \nonumber \\
+& \frac{  i \sigma_s \kappa_{ms} e^{  i  \delta_2 } (1 - \sigma_{ma} e^{ i  \delta_4}) e^{- i \tilde{\delta}}}{1- \sigma_{ma} e^{ i  \delta_4} + \sigma_{ms} e^{ i  (\delta_3 + \delta_4 + \delta_5)} (1- \sigma_{ma} e^{ - i  \delta_4})}\;. 
\label{Fielda_General_system2}
\end{align}

Finally, following the structure of \eqref{asymptotic1}, the asymptotic-in fields inside the main ring are given by composing \eqref{Fielda_General_system} and \eqref{Fielda_General_system2}, so that 
\begin{align}
    {\mathbf{D}_{(1,2),k}^\text{Ring}}(\textbf{r}_{\perp},\zeta)=&  (A_d^{(1,2)} + B_d^{(1,2)})~\frac{\mathbf{d}_{(1,2),k}(\textbf{r}_{\perp},\zeta)}{\sqrt{2\pi}} \nonumber\\ 
    = & f_{k}^{(1,2)}(\zeta) \frac{\mathbf{d}_{(1,2),k}(\textbf{r}_{\perp},\zeta)}{\sqrt{2\pi}} \;. 
    \label{asymptotic_General}
\end{align}
Also in this case we use  $\textrm{d}\textbf{r}=\textrm{d}\textbf{r}_{\perp}\textrm{d}\zeta$, where $\textbf{r}_{\perp}$ indicates the coordinates of the dipole position $R\zeta$ in the plane normal to the Sagnac, while with $\zeta$ we refer to the coordinate position in the counterclockwise direction along the ring circumference. Just to make clear the dependency of the dipole emission rate the on the phase shifts, in main text we chose to express $ \left|f_{k}^{(1,2)}(\zeta)\right|^2$ in terms of main ring phase shift $\delta_m=\delta_3+\delta_5$, aux phase shift $\delta_a=\delta_4$, and Sagnac phase shift $\delta_s=\delta_1+\delta_2$. Finally note that, due to the presence of the Sagnac interferometer, regardless of whether the photon is injected from port 1 or 2, the dipole will always see the superposition of a clockwise and a counterclockwise field. For this reason, we cannot label the asymptotic-in fields as left ($L$) and right ($R$). 

The strategy adopted to calculate the asymptotic-in/out fields can be generalized to more complex structures. For instance, it would be possible to extend the approach to multi-scatterer configurations and to multiple point-like dipole–resonator coupling sites, thereby accommodating increasing levels of complexity in integrated photonic systems. The resulting complexity amounts to computing the scattering matrix, which involves solving increasingly large systems of algebraic equations for the incoming and outgoing fields at the coupling points. Such calculations, though tedious, can be effectively addressed with numerical methods.

\newpage

\bibliography{alessandria}

\begin{thebibliography}{50}%
\makeatletter
\providecommand \@ifxundefined [1]{%
 \@ifx{#1\undefined}
}%
\providecommand \@ifnum [1]{%
 \ifnum #1\expandafter \@firstoftwo
 \else \expandafter \@secondoftwo
 \fi
}%
\providecommand \@ifx [1]{%
 \ifx #1\expandafter \@firstoftwo
 \else \expandafter \@secondoftwo
 \fi
}%
\providecommand \natexlab [1]{#1}%
\providecommand \enquote  [1]{``#1''}%
\providecommand \bibnamefont  [1]{#1}%
\providecommand \bibfnamefont [1]{#1}%
\providecommand \citenamefont [1]{#1}%
\providecommand \href@noop [0]{\@secondoftwo}%
\providecommand \href [0]{\begingroup \@sanitize@url \@href}%
\providecommand \@href[1]{\@@startlink{#1}\@@href}%
\providecommand \@@href[1]{\endgroup#1\@@endlink}%
\providecommand \@sanitize@url [0]{\catcode `\\12\catcode `\$12\catcode
  `\&12\catcode `\#12\catcode `\^12\catcode `\_12\catcode `\%12\relax}%
\providecommand \@@startlink[1]{}%
\providecommand \@@endlink[0]{}%
\providecommand \url  [0]{\begingroup\@sanitize@url \@url }%
\providecommand \@url [1]{\endgroup\@href {#1}{\urlprefix }}%
\providecommand \urlprefix  [0]{URL }%
\providecommand \Eprint [0]{\href }%
\providecommand \doibase [0]{https://doi.org/}%
\providecommand \selectlanguage [0]{\@gobble}%
\providecommand \bibinfo  [0]{\@secondoftwo}%
\providecommand \bibfield  [0]{\@secondoftwo}%
\providecommand \translation [1]{[#1]}%
\providecommand \BibitemOpen [0]{}%
\providecommand \bibitemStop [0]{}%
\providecommand \bibitemNoStop [0]{.\EOS\space}%
\providecommand \EOS [0]{\spacefactor3000\relax}%
\providecommand \BibitemShut  [1]{\csname bibitem#1\endcsname}%
\let\auto@bib@innerbib\@empty
%</preamble>
\bibitem [{\citenamefont {Fermi}(1932)}]{Fermi1932}%
  \BibitemOpen
  \bibfield  {author} {\bibinfo {author} {\bibfnamefont {E.}~\bibnamefont
  {Fermi}},\ }\href {https://doi.org/10.1103/RevModPhys.4.87} {\bibfield
  {journal} {\bibinfo  {journal} {Rev. Mod. Phys.}\ }\textbf {\bibinfo {volume}
  {4}},\ \bibinfo {pages} {87} (\bibinfo {year} {1932})}\BibitemShut {NoStop}%
\bibitem [{\citenamefont {Weisskopf}\ and\ \citenamefont
  {Wigner}(1997)}]{weisskopf1997berechnung}%
  \BibitemOpen
  \bibfield  {author} {\bibinfo {author} {\bibfnamefont {V.}~\bibnamefont
  {Weisskopf}}\ and\ \bibinfo {author} {\bibfnamefont {E.}~\bibnamefont
  {Wigner}},\ }\href
  {https://link.springer.com/chapter/10.1007/978-3-662-09203-3_3} {\bibfield
  {journal} {\bibinfo  {journal} {Part I: Particles and Fields. Part II:
  Foundations of Quantum Mechanics}\ ,\ \bibinfo {pages} {30}} (\bibinfo {year}
  {1997})}\BibitemShut {NoStop}%
\bibitem [{\citenamefont {Novotny}\ and\ \citenamefont
  {Hecht}(2012)}]{novotny2012}%
  \BibitemOpen
  \bibfield  {author} {\bibinfo {author} {\bibfnamefont {L.}~\bibnamefont
  {Novotny}}\ and\ \bibinfo {author} {\bibfnamefont {B.}~\bibnamefont
  {Hecht}},\ }\href@noop {} {\emph {\bibinfo {title} {Principles of
  nano-optics}}}\ (\bibinfo  {publisher} {Cambridge university press},\
  \bibinfo {year} {2012})\BibitemShut {NoStop}%
\bibitem [{\citenamefont {Chow}\ \emph {et~al.}(2014)\citenamefont {Chow},
  \citenamefont {Jahnke},\ and\ \citenamefont {Gies}}]{chow2014}%
  \BibitemOpen
  \bibfield  {author} {\bibinfo {author} {\bibfnamefont {W.~W.}\ \bibnamefont
  {Chow}}, \bibinfo {author} {\bibfnamefont {F.}~\bibnamefont {Jahnke}},\ and\
  \bibinfo {author} {\bibfnamefont {C.}~\bibnamefont {Gies}},\ }\href
  {https://www.nature.com/articles/lsa201482} {\bibfield  {journal} {\bibinfo
  {journal} {Light: Science \& Applications}\ }\textbf {\bibinfo {volume}
  {3}},\ \bibinfo {pages} {e201} (\bibinfo {year} {2014})}\BibitemShut
  {NoStop}%
\bibitem [{\citenamefont {Deng}\ \emph {et~al.}(2021)\citenamefont {Deng},
  \citenamefont {Lippi}, \citenamefont {M{\o}rk}, \citenamefont {Wiersig},\
  and\ \citenamefont {Reitzenstein}}]{deng2021}%
  \BibitemOpen
  \bibfield  {author} {\bibinfo {author} {\bibfnamefont {H.}~\bibnamefont
  {Deng}}, \bibinfo {author} {\bibfnamefont {G.~L.}\ \bibnamefont {Lippi}},
  \bibinfo {author} {\bibfnamefont {J.}~\bibnamefont {M{\o}rk}}, \bibinfo
  {author} {\bibfnamefont {J.}~\bibnamefont {Wiersig}},\ and\ \bibinfo {author}
  {\bibfnamefont {S.}~\bibnamefont {Reitzenstein}},\ }\href
  {https://onlinelibrary.wiley.com/doi/full/10.1002/adom.202100415} {\bibfield
  {journal} {\bibinfo  {journal} {Advanced Optical Materials}\ }\textbf
  {\bibinfo {volume} {9}},\ \bibinfo {pages} {2100415} (\bibinfo {year}
  {2021})}\BibitemShut {NoStop}%
\bibitem [{\citenamefont {S\o{}ndergaard}\ and\ \citenamefont
  {Tromborg}(2001)}]{Tromborg2001}%
  \BibitemOpen
  \bibfield  {author} {\bibinfo {author} {\bibfnamefont {T.}~\bibnamefont
  {S\o{}ndergaard}}\ and\ \bibinfo {author} {\bibfnamefont {B.}~\bibnamefont
  {Tromborg}},\ }\href {https://doi.org/10.1103/PhysRevA.64.033812} {\bibfield
  {journal} {\bibinfo  {journal} {Phys. Rev. A}\ }\textbf {\bibinfo {volume}
  {64}},\ \bibinfo {pages} {033812} (\bibinfo {year} {2001})}\BibitemShut
  {NoStop}%
\bibitem [{\citenamefont {Lin}\ \emph {et~al.}(2016)\citenamefont {Lin},
  \citenamefont {Pick}, \citenamefont {Lon\ifmmode~\check{c}\else
  \v{c}\fi{}ar},\ and\ \citenamefont {Rodriguez}}]{Lin2016}%
  \BibitemOpen
  \bibfield  {author} {\bibinfo {author} {\bibfnamefont {Z.}~\bibnamefont
  {Lin}}, \bibinfo {author} {\bibfnamefont {A.}~\bibnamefont {Pick}}, \bibinfo
  {author} {\bibfnamefont {M.}~\bibnamefont {Lon\ifmmode~\check{c}\else
  \v{c}\fi{}ar}},\ and\ \bibinfo {author} {\bibfnamefont {A.~W.}\ \bibnamefont
  {Rodriguez}},\ }\href {https://doi.org/10.1103/PhysRevLett.117.107402}
  {\bibfield  {journal} {\bibinfo  {journal} {Phys. Rev. Lett.}\ }\textbf
  {\bibinfo {volume} {117}},\ \bibinfo {pages} {107402} (\bibinfo {year}
  {2016})}\BibitemShut {NoStop}%
\bibitem [{\citenamefont {Pick}\ \emph {et~al.}(2017)\citenamefont {Pick},
  \citenamefont {Zhen}, \citenamefont {Miller}, \citenamefont {Hsu},
  \citenamefont {Hernandez}, \citenamefont {Rodriguez}, \citenamefont
  {Solja{\v{c}}i{\'c}},\ and\ \citenamefont {Johnson}}]{pick2017}%
  \BibitemOpen
  \bibfield  {author} {\bibinfo {author} {\bibfnamefont {A.}~\bibnamefont
  {Pick}}, \bibinfo {author} {\bibfnamefont {B.}~\bibnamefont {Zhen}}, \bibinfo
  {author} {\bibfnamefont {O.~D.}\ \bibnamefont {Miller}}, \bibinfo {author}
  {\bibfnamefont {C.~W.}\ \bibnamefont {Hsu}}, \bibinfo {author} {\bibfnamefont
  {F.}~\bibnamefont {Hernandez}}, \bibinfo {author} {\bibfnamefont {A.~W.}\
  \bibnamefont {Rodriguez}}, \bibinfo {author} {\bibfnamefont {M.}~\bibnamefont
  {Solja{\v{c}}i{\'c}}},\ and\ \bibinfo {author} {\bibfnamefont {S.~G.}\
  \bibnamefont {Johnson}},\ }\href
  {https://opg.optica.org/oe/fulltext.cfm?uri=oe-25-11-12325&id=366721}
  {\bibfield  {journal} {\bibinfo  {journal} {Optics express}\ }\textbf
  {\bibinfo {volume} {25}},\ \bibinfo {pages} {12325} (\bibinfo {year}
  {2017})}\BibitemShut {NoStop}%
\bibitem [{\citenamefont {Peng}\ \emph {et~al.}(2014)\citenamefont {Peng},
  \citenamefont {{\"O}zdemir}, \citenamefont {Lei}, \citenamefont {Monifi},
  \citenamefont {Gianfreda}, \citenamefont {Long}, \citenamefont {Fan},
  \citenamefont {Nori}, \citenamefont {Bender},\ and\ \citenamefont
  {Yang}}]{peng2014}%
  \BibitemOpen
  \bibfield  {author} {\bibinfo {author} {\bibfnamefont {B.}~\bibnamefont
  {Peng}}, \bibinfo {author} {\bibfnamefont {{\c{S}}.~K.}\ \bibnamefont
  {{\"O}zdemir}}, \bibinfo {author} {\bibfnamefont {F.}~\bibnamefont {Lei}},
  \bibinfo {author} {\bibfnamefont {F.}~\bibnamefont {Monifi}}, \bibinfo
  {author} {\bibfnamefont {M.}~\bibnamefont {Gianfreda}}, \bibinfo {author}
  {\bibfnamefont {G.~L.}\ \bibnamefont {Long}}, \bibinfo {author}
  {\bibfnamefont {S.}~\bibnamefont {Fan}}, \bibinfo {author} {\bibfnamefont
  {F.}~\bibnamefont {Nori}}, \bibinfo {author} {\bibfnamefont {C.~M.}\
  \bibnamefont {Bender}},\ and\ \bibinfo {author} {\bibfnamefont
  {L.}~\bibnamefont {Yang}},\ }\href
  {https://www.nature.com/articles/nphys2927} {\bibfield  {journal} {\bibinfo
  {journal} {Nature Physics}\ }\textbf {\bibinfo {volume} {10}},\ \bibinfo
  {pages} {394} (\bibinfo {year} {2014})}\BibitemShut {NoStop}%
\bibitem [{\citenamefont {Genoni}\ \emph {et~al.}(2007)\citenamefont {Genoni},
  \citenamefont {Paris},\ and\ \citenamefont {Banaszek}}]{Genoni2007}%
  \BibitemOpen
  \bibfield  {author} {\bibinfo {author} {\bibfnamefont {M.~G.}\ \bibnamefont
  {Genoni}}, \bibinfo {author} {\bibfnamefont {M.~G.~A.}\ \bibnamefont
  {Paris}},\ and\ \bibinfo {author} {\bibfnamefont {K.}~\bibnamefont
  {Banaszek}},\ }\href {https://doi.org/10.1103/PhysRevA.76.042327} {\bibfield
  {journal} {\bibinfo  {journal} {Phys. Rev. A}\ }\textbf {\bibinfo {volume}
  {76}},\ \bibinfo {pages} {042327} (\bibinfo {year} {2007})}\BibitemShut
  {NoStop}%
\bibitem [{\citenamefont {Smith}\ \emph {et~al.}(2011)\citenamefont {Smith},
  \citenamefont {Smolin},\ and\ \citenamefont {Yard}}]{smith2011}%
  \BibitemOpen
  \bibfield  {author} {\bibinfo {author} {\bibfnamefont {G.}~\bibnamefont
  {Smith}}, \bibinfo {author} {\bibfnamefont {J.~A.}\ \bibnamefont {Smolin}},\
  and\ \bibinfo {author} {\bibfnamefont {J.}~\bibnamefont {Yard}},\ }\href
  {https://www.nature.com/articles/nphoton.2011.203} {\bibfield  {journal}
  {\bibinfo  {journal} {Nature Photonics}\ }\textbf {\bibinfo {volume} {5}},\
  \bibinfo {pages} {624} (\bibinfo {year} {2011})}\BibitemShut {NoStop}%
\bibitem [{\citenamefont {Lee}\ \emph {et~al.}(2019)\citenamefont {Lee},
  \citenamefont {Park},\ and\ \citenamefont {Nha}}]{lee2019}%
  \BibitemOpen
  \bibfield  {author} {\bibinfo {author} {\bibfnamefont {J.}~\bibnamefont
  {Lee}}, \bibinfo {author} {\bibfnamefont {J.}~\bibnamefont {Park}},\ and\
  \bibinfo {author} {\bibfnamefont {H.}~\bibnamefont {Nha}},\ }\href
  {https://www.nature.com/articles/s41534-019-0164-9} {\bibfield  {journal}
  {\bibinfo  {journal} {npj Quantum Information}\ }\textbf {\bibinfo {volume}
  {5}},\ \bibinfo {pages} {49} (\bibinfo {year} {2019})}\BibitemShut {NoStop}%
\bibitem [{\citenamefont {Huang}\ \emph {et~al.}(2018)\citenamefont {Huang},
  \citenamefont {Zhuang},\ and\ \citenamefont {Lee}}]{Huang2018}%
  \BibitemOpen
  \bibfield  {author} {\bibinfo {author} {\bibfnamefont {J.}~\bibnamefont
  {Huang}}, \bibinfo {author} {\bibfnamefont {M.}~\bibnamefont {Zhuang}},\ and\
  \bibinfo {author} {\bibfnamefont {C.}~\bibnamefont {Lee}},\ }\href
  {https://doi.org/10.1103/PhysRevA.97.032116} {\bibfield  {journal} {\bibinfo
  {journal} {Phys. Rev. A}\ }\textbf {\bibinfo {volume} {97}},\ \bibinfo
  {pages} {032116} (\bibinfo {year} {2018})}\BibitemShut {NoStop}%
\bibitem [{\citenamefont {Walschaers}(2021)}]{Walschaers2021}%
  \BibitemOpen
  \bibfield  {author} {\bibinfo {author} {\bibfnamefont {M.}~\bibnamefont
  {Walschaers}},\ }\href {https://doi.org/10.1103/PRXQuantum.2.030204}
  {\bibfield  {journal} {\bibinfo  {journal} {PRX Quantum}\ }\textbf {\bibinfo
  {volume} {2}},\ \bibinfo {pages} {030204} (\bibinfo {year}
  {2021})}\BibitemShut {NoStop}%
\bibitem [{\citenamefont {Xu}\ \emph {et~al.}(2022)\citenamefont {Xu},
  \citenamefont {Zhang}, \citenamefont {Sun}, \citenamefont {Li}, \citenamefont
  {Song}, \citenamefont {Xiang}, \citenamefont {Huang}, \citenamefont {Li},
  \citenamefont {Shi}, \citenamefont {Chen}, \citenamefont {Song},
  \citenamefont {Zheng}, \citenamefont {Nori}, \citenamefont {Wang},\ and\
  \citenamefont {Fan}}]{Xu2022}%
  \BibitemOpen
  \bibfield  {author} {\bibinfo {author} {\bibfnamefont {K.}~\bibnamefont
  {Xu}}, \bibinfo {author} {\bibfnamefont {Y.-R.}\ \bibnamefont {Zhang}},
  \bibinfo {author} {\bibfnamefont {Z.-H.}\ \bibnamefont {Sun}}, \bibinfo
  {author} {\bibfnamefont {H.}~\bibnamefont {Li}}, \bibinfo {author}
  {\bibfnamefont {P.}~\bibnamefont {Song}}, \bibinfo {author} {\bibfnamefont
  {Z.}~\bibnamefont {Xiang}}, \bibinfo {author} {\bibfnamefont
  {K.}~\bibnamefont {Huang}}, \bibinfo {author} {\bibfnamefont
  {H.}~\bibnamefont {Li}}, \bibinfo {author} {\bibfnamefont {Y.-H.}\
  \bibnamefont {Shi}}, \bibinfo {author} {\bibfnamefont {C.-T.}\ \bibnamefont
  {Chen}}, \bibinfo {author} {\bibfnamefont {X.}~\bibnamefont {Song}}, \bibinfo
  {author} {\bibfnamefont {D.}~\bibnamefont {Zheng}}, \bibinfo {author}
  {\bibfnamefont {F.}~\bibnamefont {Nori}}, \bibinfo {author} {\bibfnamefont
  {H.}~\bibnamefont {Wang}},\ and\ \bibinfo {author} {\bibfnamefont
  {H.}~\bibnamefont {Fan}},\ }\href
  {https://doi.org/10.1103/PhysRevLett.128.150501} {\bibfield  {journal}
  {\bibinfo  {journal} {Phys. Rev. Lett.}\ }\textbf {\bibinfo {volume} {128}},\
  \bibinfo {pages} {150501} (\bibinfo {year} {2022})}\BibitemShut {NoStop}%
\bibitem [{\citenamefont {Senellart}\ \emph {et~al.}(2017)\citenamefont
  {Senellart}, \citenamefont {Solomon},\ and\ \citenamefont
  {White}}]{Senellart2017}%
  \BibitemOpen
  \bibfield  {author} {\bibinfo {author} {\bibfnamefont {P.}~\bibnamefont
  {Senellart}}, \bibinfo {author} {\bibfnamefont {G.}~\bibnamefont {Solomon}},\
  and\ \bibinfo {author} {\bibfnamefont {A.}~\bibnamefont {White}},\ }\href
  {https://ui.adsabs.harvard.edu/abs/2017NatNa..12.1026S/abstract} {\bibfield
  {journal} {\bibinfo  {journal} {Nature nanotechnology}\ }\textbf {\bibinfo
  {volume} {12}},\ \bibinfo {pages} {1026} (\bibinfo {year}
  {2017})}\BibitemShut {NoStop}%
\bibitem [{\citenamefont {Larocque}\ \emph {et~al.}(2024)\citenamefont
  {Larocque}, \citenamefont {Buyukkaya}, \citenamefont {Errando-Herranz},
  \citenamefont {Papon}, \citenamefont {Harper}, \citenamefont {Tao},
  \citenamefont {Carolan}, \citenamefont {Lee}, \citenamefont {Richardson},
  \citenamefont {Leake} \emph {et~al.}}]{larocque2024tunable}%
  \BibitemOpen
  \bibfield  {author} {\bibinfo {author} {\bibfnamefont {H.}~\bibnamefont
  {Larocque}}, \bibinfo {author} {\bibfnamefont {M.~A.}\ \bibnamefont
  {Buyukkaya}}, \bibinfo {author} {\bibfnamefont {C.}~\bibnamefont
  {Errando-Herranz}}, \bibinfo {author} {\bibfnamefont {C.}~\bibnamefont
  {Papon}}, \bibinfo {author} {\bibfnamefont {S.}~\bibnamefont {Harper}},
  \bibinfo {author} {\bibfnamefont {M.}~\bibnamefont {Tao}}, \bibinfo {author}
  {\bibfnamefont {J.}~\bibnamefont {Carolan}}, \bibinfo {author} {\bibfnamefont
  {C.-M.}\ \bibnamefont {Lee}}, \bibinfo {author} {\bibfnamefont {C.~J.}\
  \bibnamefont {Richardson}}, \bibinfo {author} {\bibfnamefont {G.~L.}\
  \bibnamefont {Leake}}, \emph {et~al.},\ }\href
  {https://www.nature.com/articles/s41467-024-50208-0} {\bibfield  {journal}
  {\bibinfo  {journal} {Nature Communications}\ }\textbf {\bibinfo {volume}
  {15}},\ \bibinfo {pages} {5781} (\bibinfo {year} {2024})}\BibitemShut
  {NoStop}%
\bibitem [{\citenamefont {Ruf}\ \emph {et~al.}(2019)\citenamefont {Ruf},
  \citenamefont {IJspeert}, \citenamefont {Van~Dam}, \citenamefont {De~Jong},
  \citenamefont {Van Den~Berg}, \citenamefont {Evers},\ and\ \citenamefont
  {Hanson}}]{Ruf2019}%
  \BibitemOpen
  \bibfield  {author} {\bibinfo {author} {\bibfnamefont {M.}~\bibnamefont
  {Ruf}}, \bibinfo {author} {\bibfnamefont {M.}~\bibnamefont {IJspeert}},
  \bibinfo {author} {\bibfnamefont {S.}~\bibnamefont {Van~Dam}}, \bibinfo
  {author} {\bibfnamefont {N.}~\bibnamefont {De~Jong}}, \bibinfo {author}
  {\bibfnamefont {H.}~\bibnamefont {Van Den~Berg}}, \bibinfo {author}
  {\bibfnamefont {G.}~\bibnamefont {Evers}},\ and\ \bibinfo {author}
  {\bibfnamefont {R.}~\bibnamefont {Hanson}},\ }\href
  {https://ui.adsabs.harvard.edu/abs/2017NatNa..12.1026S/abstract} {\bibfield
  {journal} {\bibinfo  {journal} {Nano letters}\ }\textbf {\bibinfo {volume}
  {19}},\ \bibinfo {pages} {3987} (\bibinfo {year} {2019})}\BibitemShut
  {NoStop}%
\bibitem [{\citenamefont {Groiseau}\ \emph {et~al.}(2024)\citenamefont
  {Groiseau}, \citenamefont {Fern\'andez-Dom\'{\i}nguez}, \citenamefont
  {Mart\'{\i}n-Cano},\ and\ \citenamefont {Mu\~noz}}]{Groiseau2024}%
  \BibitemOpen
  \bibfield  {author} {\bibinfo {author} {\bibfnamefont {C.}~\bibnamefont
  {Groiseau}}, \bibinfo {author} {\bibfnamefont {A.~I.}\ \bibnamefont
  {Fern\'andez-Dom\'{\i}nguez}}, \bibinfo {author} {\bibfnamefont
  {D.}~\bibnamefont {Mart\'{\i}n-Cano}},\ and\ \bibinfo {author} {\bibfnamefont
  {C.~S.}\ \bibnamefont {Mu\~noz}},\ }\href
  {https://doi.org/10.1103/PRXQuantum.5.010312} {\bibfield  {journal} {\bibinfo
   {journal} {PRX Quantum}\ }\textbf {\bibinfo {volume} {5}},\ \bibinfo {pages}
  {010312} (\bibinfo {year} {2024})}\BibitemShut {NoStop}%
\bibitem [{\citenamefont {Liscidini}\ \emph {et~al.}(2012)\citenamefont
  {Liscidini}, \citenamefont {Helt},\ and\ \citenamefont
  {Sipe}}]{Liscidini2012}%
  \BibitemOpen
  \bibfield  {author} {\bibinfo {author} {\bibfnamefont {M.}~\bibnamefont
  {Liscidini}}, \bibinfo {author} {\bibfnamefont {L.~G.}\ \bibnamefont
  {Helt}},\ and\ \bibinfo {author} {\bibfnamefont {J.~E.}\ \bibnamefont
  {Sipe}},\ }\href {https://doi.org/10.1103/PhysRevA.85.013833} {\bibfield
  {journal} {\bibinfo  {journal} {Phys. Rev. A}\ }\textbf {\bibinfo {volume}
  {85}},\ \bibinfo {pages} {013833} (\bibinfo {year} {2012})}\BibitemShut
  {NoStop}%
\bibitem [{\citenamefont {Dirac}(1927)}]{dirac1927}%
  \BibitemOpen
  \bibfield  {author} {\bibinfo {author} {\bibfnamefont {P.~A.~M.}\
  \bibnamefont {Dirac}},\ }\href
  {https://royalsocietypublishing.org/doi/abs/10.1098/rspa.1927.0039}
  {\bibfield  {journal} {\bibinfo  {journal} {Proceedings of the Royal Society
  of London. Series A, Containing Papers of a Mathematical and Physical
  Character}\ }\textbf {\bibinfo {volume} {114}},\ \bibinfo {pages} {243}
  (\bibinfo {year} {1927})}\BibitemShut {NoStop}%
\bibitem [{\citenamefont {Fermi}(1950)}]{fermi1950}%
  \BibitemOpen
  \bibfield  {author} {\bibinfo {author} {\bibfnamefont {E.}~\bibnamefont
  {Fermi}},\ }\href@noop {} {\emph {\bibinfo {title} {Nuclear physics: a course
  given by Enrico Fermi at the University of Chicago}}}\ (\bibinfo  {publisher}
  {University of Chicago press},\ \bibinfo {year} {1950})\BibitemShut {NoStop}%
\bibitem [{\citenamefont {Sakurai}\ and\ \citenamefont
  {Napolitano}(2020)}]{sakurai2020}%
  \BibitemOpen
  \bibfield  {author} {\bibinfo {author} {\bibfnamefont {J.~J.}\ \bibnamefont
  {Sakurai}}\ and\ \bibinfo {author} {\bibfnamefont {J.}~\bibnamefont
  {Napolitano}},\ }\href@noop {} {\emph {\bibinfo {title} {Modern quantum
  mechanics}}}\ (\bibinfo  {publisher} {Cambridge University Press},\ \bibinfo
  {year} {2020})\BibitemShut {NoStop}%
\bibitem [{\citenamefont {Quesada}\ \emph {et~al.}(2022)\citenamefont
  {Quesada}, \citenamefont {Helt}, \citenamefont {Menotti}, \citenamefont
  {Liscidini},\ and\ \citenamefont {Sipe}}]{quesada2022}%
  \BibitemOpen
  \bibfield  {author} {\bibinfo {author} {\bibfnamefont {N.}~\bibnamefont
  {Quesada}}, \bibinfo {author} {\bibfnamefont {L.}~\bibnamefont {Helt}},
  \bibinfo {author} {\bibfnamefont {M.}~\bibnamefont {Menotti}}, \bibinfo
  {author} {\bibfnamefont {M.}~\bibnamefont {Liscidini}},\ and\ \bibinfo
  {author} {\bibfnamefont {J.}~\bibnamefont {Sipe}},\ }\href
  {https://opg.optica.org/aop/fulltext.cfm?uri=aop-14-3-291&id=477855}
  {\bibfield  {journal} {\bibinfo  {journal} {Advances in Optics and
  Photonics}\ }\textbf {\bibinfo {volume} {14}},\ \bibinfo {pages} {291}
  (\bibinfo {year} {2022})}\BibitemShut {NoStop}%
\bibitem [{\citenamefont {Yang}\ \emph {et~al.}(2008)\citenamefont {Yang},
  \citenamefont {Liscidini},\ and\ \citenamefont {Sipe}}]{Yang2008}%
  \BibitemOpen
  \bibfield  {author} {\bibinfo {author} {\bibfnamefont {Z.}~\bibnamefont
  {Yang}}, \bibinfo {author} {\bibfnamefont {M.}~\bibnamefont {Liscidini}},\
  and\ \bibinfo {author} {\bibfnamefont {J.~E.}\ \bibnamefont {Sipe}},\ }\href
  {https://doi.org/10.1103/PhysRevA.77.033808} {\bibfield  {journal} {\bibinfo
  {journal} {Phys. Rev. A}\ }\textbf {\bibinfo {volume} {77}},\ \bibinfo
  {pages} {033808} (\bibinfo {year} {2008})}\BibitemShut {NoStop}%
\bibitem [{\citenamefont {Le~Kien}\ and\ \citenamefont
  {Rauschenbeutel}(2016)}]{Kien2016}%
  \BibitemOpen
  \bibfield  {author} {\bibinfo {author} {\bibfnamefont {F.}~\bibnamefont
  {Le~Kien}}\ and\ \bibinfo {author} {\bibfnamefont {A.}~\bibnamefont
  {Rauschenbeutel}},\ }\href {https://doi.org/10.1103/PhysRevA.93.043828}
  {\bibfield  {journal} {\bibinfo  {journal} {Phys. Rev. A}\ }\textbf {\bibinfo
  {volume} {93}},\ \bibinfo {pages} {043828} (\bibinfo {year}
  {2016})}\BibitemShut {NoStop}%
\bibitem [{\citenamefont {Bogaerts}\ \emph {et~al.}(2012)\citenamefont
  {Bogaerts}, \citenamefont {De~Heyn}, \citenamefont {Van~Vaerenbergh},
  \citenamefont {De~Vos}, \citenamefont {Kumar~Selvaraja}, \citenamefont
  {Claes}, \citenamefont {Dumon}, \citenamefont {Bienstman}, \citenamefont
  {Van~Thourhout},\ and\ \citenamefont {Baets}}]{bogaerts2012}%
  \BibitemOpen
  \bibfield  {author} {\bibinfo {author} {\bibfnamefont {W.}~\bibnamefont
  {Bogaerts}}, \bibinfo {author} {\bibfnamefont {P.}~\bibnamefont {De~Heyn}},
  \bibinfo {author} {\bibfnamefont {T.}~\bibnamefont {Van~Vaerenbergh}},
  \bibinfo {author} {\bibfnamefont {K.}~\bibnamefont {De~Vos}}, \bibinfo
  {author} {\bibfnamefont {S.}~\bibnamefont {Kumar~Selvaraja}}, \bibinfo
  {author} {\bibfnamefont {T.}~\bibnamefont {Claes}}, \bibinfo {author}
  {\bibfnamefont {P.}~\bibnamefont {Dumon}}, \bibinfo {author} {\bibfnamefont
  {P.}~\bibnamefont {Bienstman}}, \bibinfo {author} {\bibfnamefont
  {D.}~\bibnamefont {Van~Thourhout}},\ and\ \bibinfo {author} {\bibfnamefont
  {R.}~\bibnamefont {Baets}},\ }\href
  {https://onlinelibrary.wiley.com/doi/abs/10.1002/lpor.201100017?casa_token=kOI_cdeNzCMAAAAA:S1rIK75Op_Y1Tb9l5SZRaNEeNovoTleAQhf55xqamN4rFIf3fLTbZXS1Px4ZyrcdcFgxmLUsdVSGujE}
  {\bibfield  {journal} {\bibinfo  {journal} {Laser \& Photonics Reviews}\
  }\textbf {\bibinfo {volume} {6}},\ \bibinfo {pages} {47} (\bibinfo {year}
  {2012})}\BibitemShut {NoStop}%
\bibitem [{\citenamefont {O'Faolain}\ \emph {et~al.}(2010)\citenamefont
  {O'Faolain}, \citenamefont {Schulz}, \citenamefont {Beggs}, \citenamefont
  {White}, \citenamefont {Spasenovi\'{c}}, \citenamefont {Kuipers},
  \citenamefont {Morichetti}, \citenamefont {Melloni}, \citenamefont {Mazoyer},
  \citenamefont {Hugonin}, \citenamefont {Lalanne},\ and\ \citenamefont
  {Krauss}}]{OFaolain:10}%
  \BibitemOpen
  \bibfield  {author} {\bibinfo {author} {\bibfnamefont {L.}~\bibnamefont
  {O'Faolain}}, \bibinfo {author} {\bibfnamefont {S.~A.}\ \bibnamefont
  {Schulz}}, \bibinfo {author} {\bibfnamefont {D.~M.}\ \bibnamefont {Beggs}},
  \bibinfo {author} {\bibfnamefont {T.~P.}\ \bibnamefont {White}}, \bibinfo
  {author} {\bibfnamefont {M.}~\bibnamefont {Spasenovi\'{c}}}, \bibinfo
  {author} {\bibfnamefont {L.}~\bibnamefont {Kuipers}}, \bibinfo {author}
  {\bibfnamefont {F.}~\bibnamefont {Morichetti}}, \bibinfo {author}
  {\bibfnamefont {A.}~\bibnamefont {Melloni}}, \bibinfo {author} {\bibfnamefont
  {S.}~\bibnamefont {Mazoyer}}, \bibinfo {author} {\bibfnamefont {J.~P.}\
  \bibnamefont {Hugonin}}, \bibinfo {author} {\bibfnamefont {P.}~\bibnamefont
  {Lalanne}},\ and\ \bibinfo {author} {\bibfnamefont {T.~F.}\ \bibnamefont
  {Krauss}},\ }\href {https://doi.org/10.1364/OE.18.027627} {\bibfield
  {journal} {\bibinfo  {journal} {Opt. Express}\ }\textbf {\bibinfo {volume}
  {18}},\ \bibinfo {pages} {27627} (\bibinfo {year} {2010})}\BibitemShut
  {NoStop}%
\bibitem [{\citenamefont {Purcell}(1995)}]{purcell1995}%
  \BibitemOpen
  \bibfield  {author} {\bibinfo {author} {\bibfnamefont {E.~M.}\ \bibnamefont
  {Purcell}},\ }in\ \href
  {https://link.springer.com/chapter/10.1007/978-1-4615-1963-8_40} {\emph
  {\bibinfo {booktitle} {Confined electrons and photons: new physics and
  applications}}}\ (\bibinfo  {publisher} {Springer},\ \bibinfo {year} {1995})\
  pp.\ \bibinfo {pages} {839--839}\BibitemShut {NoStop}%
\bibitem [{\citenamefont {Onodera}\ \emph {et~al.}(2016)\citenamefont
  {Onodera}, \citenamefont {Liscidini}, \citenamefont {Sipe},\ and\
  \citenamefont {Helt}}]{onodera2016}%
  \BibitemOpen
  \bibfield  {author} {\bibinfo {author} {\bibfnamefont {T.}~\bibnamefont
  {Onodera}}, \bibinfo {author} {\bibfnamefont {M.}~\bibnamefont {Liscidini}},
  \bibinfo {author} {\bibfnamefont {J.~E.}\ \bibnamefont {Sipe}},\ and\
  \bibinfo {author} {\bibfnamefont {L.~G.}\ \bibnamefont {Helt}},\ }\href
  {https://doi.org/10.1103/PhysRevA.93.043837} {\bibfield  {journal} {\bibinfo
  {journal} {Phys. Rev. A}\ }\textbf {\bibinfo {volume} {93}},\ \bibinfo
  {pages} {043837} (\bibinfo {year} {2016})}\BibitemShut {NoStop}%
\bibitem [{\citenamefont {Rosenfeld}(2003)}]{rosenfeld2003}%
  \BibitemOpen
  \bibfield  {author} {\bibinfo {author} {\bibfnamefont {W.}~\bibnamefont
  {Rosenfeld}},\ }\href
  {https://quantum-technologies.iap.uni-bonn.de/de/diplom-theses.html?task=download&file=78&token=1c344fe05720a619499ada4ffe4f3814}
  {\bibfield  {journal} {\bibinfo  {journal} {Diplom thesis, Universit{\"a}t
  Bonn}\ } (\bibinfo {year} {2003})}\BibitemShut {NoStop}%
\bibitem [{\citenamefont {Fait}\ \emph {et~al.}(2021)\citenamefont {Fait},
  \citenamefont {Putz}, \citenamefont {Wachter}, \citenamefont {Schalko},
  \citenamefont {Schmid}, \citenamefont {Arndt},\ and\ \citenamefont
  {Trupke}}]{fait2021}%
  \BibitemOpen
  \bibfield  {author} {\bibinfo {author} {\bibfnamefont {J.}~\bibnamefont
  {Fait}}, \bibinfo {author} {\bibfnamefont {S.}~\bibnamefont {Putz}}, \bibinfo
  {author} {\bibfnamefont {G.}~\bibnamefont {Wachter}}, \bibinfo {author}
  {\bibfnamefont {J.}~\bibnamefont {Schalko}}, \bibinfo {author} {\bibfnamefont
  {U.}~\bibnamefont {Schmid}}, \bibinfo {author} {\bibfnamefont
  {M.}~\bibnamefont {Arndt}},\ and\ \bibinfo {author} {\bibfnamefont
  {M.}~\bibnamefont {Trupke}},\ }\href
  {https://pubs.aip.org/aip/apl/article/119/22/221112/40930} {\bibfield
  {journal} {\bibinfo  {journal} {Applied Physics Letters}\ }\textbf {\bibinfo
  {volume} {119}} (\bibinfo {year} {2021})}\BibitemShut {NoStop}%
\bibitem [{\citenamefont {Fl{\aa}gan}\ \emph {et~al.}(2022)\citenamefont
  {Fl{\aa}gan}, \citenamefont {Riedel}, \citenamefont {Javadi}, \citenamefont
  {Jakubczyk}, \citenamefont {Maletinsky},\ and\ \citenamefont
  {Warburton}}]{flaagan2022}%
  \BibitemOpen
  \bibfield  {author} {\bibinfo {author} {\bibfnamefont {S.}~\bibnamefont
  {Fl{\aa}gan}}, \bibinfo {author} {\bibfnamefont {D.}~\bibnamefont {Riedel}},
  \bibinfo {author} {\bibfnamefont {A.}~\bibnamefont {Javadi}}, \bibinfo
  {author} {\bibfnamefont {T.}~\bibnamefont {Jakubczyk}}, \bibinfo {author}
  {\bibfnamefont {P.}~\bibnamefont {Maletinsky}},\ and\ \bibinfo {author}
  {\bibfnamefont {R.~J.}\ \bibnamefont {Warburton}},\ }\href
  {https://pubs.aip.org/aip/jap/article/131/11/113102/2836604} {\bibfield
  {journal} {\bibinfo  {journal} {Journal of Applied Physics}\ }\textbf
  {\bibinfo {volume} {131}} (\bibinfo {year} {2022})}\BibitemShut {NoStop}%
\bibitem [{\citenamefont {Poulton}\ \emph {et~al.}(2006)\citenamefont
  {Poulton}, \citenamefont {Koos}, \citenamefont {Fujii}, \citenamefont
  {Pfrang}, \citenamefont {Schimmel}, \citenamefont {Leuthold},\ and\
  \citenamefont {Freude}}]{poulton2006}%
  \BibitemOpen
  \bibfield  {author} {\bibinfo {author} {\bibfnamefont {C.~G.}\ \bibnamefont
  {Poulton}}, \bibinfo {author} {\bibfnamefont {C.}~\bibnamefont {Koos}},
  \bibinfo {author} {\bibfnamefont {M.}~\bibnamefont {Fujii}}, \bibinfo
  {author} {\bibfnamefont {A.}~\bibnamefont {Pfrang}}, \bibinfo {author}
  {\bibfnamefont {T.}~\bibnamefont {Schimmel}}, \bibinfo {author}
  {\bibfnamefont {J.}~\bibnamefont {Leuthold}},\ and\ \bibinfo {author}
  {\bibfnamefont {W.}~\bibnamefont {Freude}},\ }\href
  {https://ieeexplore.ieee.org/abstract/document/4032688?casa_token=hs8FFXfNmcgAAAAA:mEky00hftrLDXMCJJAS5S1IDVB6g7Ihpi0SE-omIza_xg7icLhTahai1ncLj47tmAapk03EHcA}
  {\bibfield  {journal} {\bibinfo  {journal} {IEEE Journal of selected topics
  in quantum electronics}\ }\textbf {\bibinfo {volume} {12}},\ \bibinfo {pages}
  {1306} (\bibinfo {year} {2006})}\BibitemShut {NoStop}%
\bibitem [{\citenamefont {Morichetti}\ \emph {et~al.}(2010)\citenamefont
  {Morichetti}, \citenamefont {Canciamilla}, \citenamefont {Ferrari},
  \citenamefont {Torregiani}, \citenamefont {Melloni},\ and\ \citenamefont
  {Martinelli}}]{Morichetti2010}%
  \BibitemOpen
  \bibfield  {author} {\bibinfo {author} {\bibfnamefont {F.}~\bibnamefont
  {Morichetti}}, \bibinfo {author} {\bibfnamefont {A.}~\bibnamefont
  {Canciamilla}}, \bibinfo {author} {\bibfnamefont {C.}~\bibnamefont
  {Ferrari}}, \bibinfo {author} {\bibfnamefont {M.}~\bibnamefont {Torregiani}},
  \bibinfo {author} {\bibfnamefont {A.}~\bibnamefont {Melloni}},\ and\ \bibinfo
  {author} {\bibfnamefont {M.}~\bibnamefont {Martinelli}},\ }\href
  {https://doi.org/10.1103/PhysRevLett.104.033902} {\bibfield  {journal}
  {\bibinfo  {journal} {Phys. Rev. Lett.}\ }\textbf {\bibinfo {volume} {104}},\
  \bibinfo {pages} {033902} (\bibinfo {year} {2010})}\BibitemShut {NoStop}%
\bibitem [{\citenamefont {Lee}\ \emph {et~al.}(2001)\citenamefont {Lee},
  \citenamefont {Lim}, \citenamefont {Kimerling}, \citenamefont {Shin},\ and\
  \citenamefont {Cerrina}}]{lee2001}%
  \BibitemOpen
  \bibfield  {author} {\bibinfo {author} {\bibfnamefont {K.~K.}\ \bibnamefont
  {Lee}}, \bibinfo {author} {\bibfnamefont {D.~R.}\ \bibnamefont {Lim}},
  \bibinfo {author} {\bibfnamefont {L.~C.}\ \bibnamefont {Kimerling}}, \bibinfo
  {author} {\bibfnamefont {J.}~\bibnamefont {Shin}},\ and\ \bibinfo {author}
  {\bibfnamefont {F.}~\bibnamefont {Cerrina}},\ }\href
  {https://opg.optica.org/ol/abstract.cfm?uri=ol-26-23-1888} {\bibfield
  {journal} {\bibinfo  {journal} {Optics letters}\ }\textbf {\bibinfo {volume}
  {26}},\ \bibinfo {pages} {1888} (\bibinfo {year} {2001})}\BibitemShut
  {NoStop}%
\bibitem [{\citenamefont {Smith}\ \emph {et~al.}(2000)\citenamefont {Smith},
  \citenamefont {Benisty}, \citenamefont {Olivier}, \citenamefont {Rattier},
  \citenamefont {Weisbuch}, \citenamefont {Krauss}, \citenamefont {De~La~Rue},
  \citenamefont {Houdr{\'e}},\ and\ \citenamefont {Oesterle}}]{smith2000}%
  \BibitemOpen
  \bibfield  {author} {\bibinfo {author} {\bibfnamefont {C.}~\bibnamefont
  {Smith}}, \bibinfo {author} {\bibfnamefont {H.}~\bibnamefont {Benisty}},
  \bibinfo {author} {\bibfnamefont {S.}~\bibnamefont {Olivier}}, \bibinfo
  {author} {\bibfnamefont {M.}~\bibnamefont {Rattier}}, \bibinfo {author}
  {\bibfnamefont {C.}~\bibnamefont {Weisbuch}}, \bibinfo {author}
  {\bibfnamefont {T.}~\bibnamefont {Krauss}}, \bibinfo {author} {\bibfnamefont
  {R.}~\bibnamefont {De~La~Rue}}, \bibinfo {author} {\bibfnamefont
  {R.}~\bibnamefont {Houdr{\'e}}},\ and\ \bibinfo {author} {\bibfnamefont
  {U.}~\bibnamefont {Oesterle}},\ }\href
  {https://pubs.aip.org/aip/apl/article-abstract/77/18/2813/148531/Low-loss-channel-waveguides-with-two-dimensional}
  {\bibfield  {journal} {\bibinfo  {journal} {Applied Physics Letters}\
  }\textbf {\bibinfo {volume} {77}},\ \bibinfo {pages} {2813} (\bibinfo {year}
  {2000})}\BibitemShut {NoStop}%
\bibitem [{\citenamefont {Hughes}\ \emph {et~al.}(2005)\citenamefont {Hughes},
  \citenamefont {Ramunno}, \citenamefont {Young},\ and\ \citenamefont
  {Sipe}}]{Hughes2005}%
  \BibitemOpen
  \bibfield  {author} {\bibinfo {author} {\bibfnamefont {S.}~\bibnamefont
  {Hughes}}, \bibinfo {author} {\bibfnamefont {L.}~\bibnamefont {Ramunno}},
  \bibinfo {author} {\bibfnamefont {J.~F.}\ \bibnamefont {Young}},\ and\
  \bibinfo {author} {\bibfnamefont {J.~E.}\ \bibnamefont {Sipe}},\ }\href
  {https://doi.org/10.1103/PhysRevLett.94.033903} {\bibfield  {journal}
  {\bibinfo  {journal} {Phys. Rev. Lett.}\ }\textbf {\bibinfo {volume} {94}},\
  \bibinfo {pages} {033903} (\bibinfo {year} {2005})}\BibitemShut {NoStop}%
\bibitem [{\citenamefont {Li}\ \emph {et~al.}(2016)\citenamefont {Li},
  \citenamefont {Van~Vaerenbergh}, \citenamefont {De~Heyn}, \citenamefont
  {Bienstman},\ and\ \citenamefont {Bogaerts}}]{li2016}%
  \BibitemOpen
  \bibfield  {author} {\bibinfo {author} {\bibfnamefont {A.}~\bibnamefont
  {Li}}, \bibinfo {author} {\bibfnamefont {T.}~\bibnamefont {Van~Vaerenbergh}},
  \bibinfo {author} {\bibfnamefont {P.}~\bibnamefont {De~Heyn}}, \bibinfo
  {author} {\bibfnamefont {P.}~\bibnamefont {Bienstman}},\ and\ \bibinfo
  {author} {\bibfnamefont {W.}~\bibnamefont {Bogaerts}},\ }\href
  {https://onlinelibrary.wiley.com/doi/full/10.1002/lpor.201500207?casa_token=p7oYAjKSLOAAAAAA%3AjA_wH5XdNaVxugZWrlOSwZFCg5ddMxYKfXZDriqJZ3apS5ycZq7P98FV5KRpoGL0LcV2zf03_OY1_M0}
  {\bibfield  {journal} {\bibinfo  {journal} {Laser \& Photonics Reviews}\
  }\textbf {\bibinfo {volume} {10}},\ \bibinfo {pages} {420} (\bibinfo {year}
  {2016})}\BibitemShut {NoStop}%
\bibitem [{\citenamefont {Little}\ \emph {et~al.}(1997)\citenamefont {Little},
  \citenamefont {Laine},\ and\ \citenamefont {Chu}}]{Little_1997}%
  \BibitemOpen
  \bibfield  {author} {\bibinfo {author} {\bibfnamefont {B.~E.}\ \bibnamefont
  {Little}}, \bibinfo {author} {\bibfnamefont {J.-P.}\ \bibnamefont {Laine}},\
  and\ \bibinfo {author} {\bibfnamefont {S.~T.}\ \bibnamefont {Chu}},\ }\href
  {https://doi.org/10.1364/OL.22.000004} {\bibfield  {journal} {\bibinfo
  {journal} {Opt. Lett.}\ }\textbf {\bibinfo {volume} {22}},\ \bibinfo {pages}
  {4} (\bibinfo {year} {1997})}\BibitemShut {NoStop}%
\bibitem [{\citenamefont {Arbabi}\ \emph {et~al.}(2011)\citenamefont {Arbabi},
  \citenamefont {Kang}, \citenamefont {Lu}, \citenamefont {Chow},\ and\
  \citenamefont {Goddard}}]{arabi_2011}%
  \BibitemOpen
  \bibfield  {author} {\bibinfo {author} {\bibfnamefont {A.}~\bibnamefont
  {Arbabi}}, \bibinfo {author} {\bibfnamefont {Y.~M.}\ \bibnamefont {Kang}},
  \bibinfo {author} {\bibfnamefont {C.-Y.}\ \bibnamefont {Lu}}, \bibinfo
  {author} {\bibfnamefont {E.}~\bibnamefont {Chow}},\ and\ \bibinfo {author}
  {\bibfnamefont {L.~L.}\ \bibnamefont {Goddard}},\ }\href
  {https://doi.org/10.1063/1.3633111} {\bibfield  {journal} {\bibinfo
  {journal} {Applied Physics Letters}\ }\textbf {\bibinfo {volume} {99}},\
  \bibinfo {pages} {091105} (\bibinfo {year} {2011})}\BibitemShut {NoStop}%
\bibitem [{\citenamefont {Tarasenko}\ and\ \citenamefont
  {Margulis}(2007)}]{tarasenko2007}%
  \BibitemOpen
  \bibfield  {author} {\bibinfo {author} {\bibfnamefont {O.}~\bibnamefont
  {Tarasenko}}\ and\ \bibinfo {author} {\bibfnamefont {W.}~\bibnamefont
  {Margulis}},\ }\href
  {https://opg.optica.org/ol/abstract.cfm?uri=ol-32-11-1356} {\bibfield
  {journal} {\bibinfo  {journal} {Optics letters}\ }\textbf {\bibinfo {volume}
  {32}},\ \bibinfo {pages} {1356} (\bibinfo {year} {2007})}\BibitemShut
  {NoStop}%
\bibitem [{\citenamefont {Liu}\ \emph {et~al.}(2015)\citenamefont {Liu},
  \citenamefont {Ye}, \citenamefont {Khan},\ and\ \citenamefont
  {Sorger}}]{liu2015}%
  \BibitemOpen
  \bibfield  {author} {\bibinfo {author} {\bibfnamefont {K.}~\bibnamefont
  {Liu}}, \bibinfo {author} {\bibfnamefont {C.~R.}\ \bibnamefont {Ye}},
  \bibinfo {author} {\bibfnamefont {S.}~\bibnamefont {Khan}},\ and\ \bibinfo
  {author} {\bibfnamefont {V.~J.}\ \bibnamefont {Sorger}},\ }\href
  {https://onlinelibrary.wiley.com/doi/full/10.1002/lpor.201400219?casa_token=OQ6Re_3ayVYAAAAA%3ADhCUYI8vlT4Ymp9xeECuBLJYflAA4EcbYv_7zU3eQ4Q6kylR5bRqkMbUv35LAitMJBGPvtvVlFybYA}
  {\bibfield  {journal} {\bibinfo  {journal} {Laser \& Photonics Reviews}\
  }\textbf {\bibinfo {volume} {9}},\ \bibinfo {pages} {172} (\bibinfo {year}
  {2015})}\BibitemShut {NoStop}%
\bibitem [{\citenamefont {Sinatkas}\ \emph {et~al.}(2021)\citenamefont
  {Sinatkas}, \citenamefont {Christopoulos}, \citenamefont {Tsilipakos},\ and\
  \citenamefont {Kriezis}}]{sinatkas2021}%
  \BibitemOpen
  \bibfield  {author} {\bibinfo {author} {\bibfnamefont {G.}~\bibnamefont
  {Sinatkas}}, \bibinfo {author} {\bibfnamefont {T.}~\bibnamefont
  {Christopoulos}}, \bibinfo {author} {\bibfnamefont {O.}~\bibnamefont
  {Tsilipakos}},\ and\ \bibinfo {author} {\bibfnamefont {E.~E.}\ \bibnamefont
  {Kriezis}},\ }\href
  {https://pubs.aip.org/aip/jap/article/130/1/010901/158529} {\bibfield
  {journal} {\bibinfo  {journal} {Journal of Applied Physics}\ }\textbf
  {\bibinfo {volume} {130}} (\bibinfo {year} {2021})}\BibitemShut {NoStop}%
\bibitem [{\citenamefont {Banic}\ \emph {et~al.}(2022)\citenamefont {Banic},
  \citenamefont {Zatti}, \citenamefont {Liscidini},\ and\ \citenamefont
  {Sipe}}]{banic22}%
  \BibitemOpen
  \bibfield  {author} {\bibinfo {author} {\bibfnamefont {M.}~\bibnamefont
  {Banic}}, \bibinfo {author} {\bibfnamefont {L.}~\bibnamefont {Zatti}},
  \bibinfo {author} {\bibfnamefont {M.}~\bibnamefont {Liscidini}},\ and\
  \bibinfo {author} {\bibfnamefont {J.~E.}\ \bibnamefont {Sipe}},\ }\href
  {https://doi.org/10.1103/PhysRevA.106.043707} {\bibfield  {journal} {\bibinfo
   {journal} {Phys. Rev. A}\ }\textbf {\bibinfo {volume} {106}},\ \bibinfo
  {pages} {043707} (\bibinfo {year} {2022})}\BibitemShut {NoStop}%
\bibitem [{\citenamefont {Simmons}(2024)}]{Simmons2024}%
  \BibitemOpen
  \bibfield  {author} {\bibinfo {author} {\bibfnamefont {S.}~\bibnamefont
  {Simmons}},\ }\href {https://doi.org/10.1103/PRXQuantum.5.010102} {\bibfield
  {journal} {\bibinfo  {journal} {PRX Quantum}\ }\textbf {\bibinfo {volume}
  {5}},\ \bibinfo {pages} {010102} (\bibinfo {year} {2024})}\BibitemShut
  {NoStop}%
\bibitem [{\citenamefont {Flamini}\ \emph {et~al.}(2018)\citenamefont
  {Flamini}, \citenamefont {Spagnolo},\ and\ \citenamefont
  {Sciarrino}}]{flamini2018photonic}%
  \BibitemOpen
  \bibfield  {author} {\bibinfo {author} {\bibfnamefont {F.}~\bibnamefont
  {Flamini}}, \bibinfo {author} {\bibfnamefont {N.}~\bibnamefont {Spagnolo}},\
  and\ \bibinfo {author} {\bibfnamefont {F.}~\bibnamefont {Sciarrino}},\ }\href
  {https://iopscience.iop.org/article/10.1088/1361-6633/aad5b2/meta?casa_token=lZxfExDIUZ4AAAAA:VdGkPZzaYdm24_xkBKLasa8hRmwDM3cPdJYqKdU9ZZNyQIJsHEdNzVc9fZ_0uS8MSLfYFL_CwS5t_1MCaHV8vNBQFA}
  {\bibfield  {journal} {\bibinfo  {journal} {Reports on Progress in Physics}\
  }\textbf {\bibinfo {volume} {82}},\ \bibinfo {pages} {016001} (\bibinfo
  {year} {2018})}\BibitemShut {NoStop}%
\bibitem [{\citenamefont {Bennett}\ and\ \citenamefont
  {Brassard}(2014)}]{bennett2014quantum}%
  \BibitemOpen
  \bibfield  {author} {\bibinfo {author} {\bibfnamefont {C.~H.}\ \bibnamefont
  {Bennett}}\ and\ \bibinfo {author} {\bibfnamefont {G.}~\bibnamefont
  {Brassard}},\ }\href
  {https://www.sciencedirect.com/science/article/pii/S0304397514004241}
  {\bibfield  {journal} {\bibinfo  {journal} {Theoretical computer science}\
  }\textbf {\bibinfo {volume} {560}},\ \bibinfo {pages} {7} (\bibinfo {year}
  {2014})}\BibitemShut {NoStop}%
\bibitem [{\citenamefont {Zhang}\ and\ \citenamefont
  {Zhuang}(2021)}]{zhang2021distributed}%
  \BibitemOpen
  \bibfield  {author} {\bibinfo {author} {\bibfnamefont {Z.}~\bibnamefont
  {Zhang}}\ and\ \bibinfo {author} {\bibfnamefont {Q.}~\bibnamefont {Zhuang}},\
  }\href
  {https://iopscience.iop.org/article/10.1088/2058-9565/abd4c3/meta?casa_token=LWXQMWkO-H8AAAAA:PY3O3unLAXRzpwQREgKDQ_RJnUN6gLFHK5kFo-iXjlsgx0r7m5XEXSioKZ2zPGNSFNnIyTTzWo8AdC3zLDaMFaskwA}
  {\bibfield  {journal} {\bibinfo  {journal} {Quantum Science and Technology}\
  }\textbf {\bibinfo {volume} {6}},\ \bibinfo {pages} {043001} (\bibinfo {year}
  {2021})}\BibitemShut {NoStop}%
\bibitem [{\citenamefont {Sipe}\ \emph {et~al.}(2004)\citenamefont {Sipe},
  \citenamefont {Bhat}, \citenamefont {Chak},\ and\ \citenamefont
  {Pereira}}]{Sipe2004}%
  \BibitemOpen
  \bibfield  {author} {\bibinfo {author} {\bibfnamefont {J.~E.}\ \bibnamefont
  {Sipe}}, \bibinfo {author} {\bibfnamefont {N.~A.~R.}\ \bibnamefont {Bhat}},
  \bibinfo {author} {\bibfnamefont {P.}~\bibnamefont {Chak}},\ and\ \bibinfo
  {author} {\bibfnamefont {S.}~\bibnamefont {Pereira}},\ }\href
  {https://doi.org/10.1103/PhysRevE.69.016604} {\bibfield  {journal} {\bibinfo
  {journal} {Phys. Rev. E}\ }\textbf {\bibinfo {volume} {69}},\ \bibinfo
  {pages} {016604} (\bibinfo {year} {2004})}\BibitemShut {NoStop}%
\end{thebibliography}%

\end{document}